\documentclass[10pt, conference, letterpaper]{IEEEtran}
\usepackage{times,color,amsmath,amssymb,epsfig,subfigure,comment,graphicx,algorithm,algorithmic,cite,multirow,enumerate}
\usepackage{balance}  
\usepackage{url}

\newtheorem{example}{Example}
\newtheorem{definition}{Definition}

\newcommand{\ppodcopt}{PPODC$_{\textrm{opt}}$}

\newcommand{\myfloor}[2]{\left \lfloor \frac{#1}{#2} \right \rfloor}

\begin{document}

\title{Privacy-Preserving and Outsourced Multi-User {\em k}-Means Clustering}


\author{%
{Bharath K. Samanthula$^\dag$, Fang-Yu Rao$^\dag$, Elisa Bertino$^\dag$, Xun Yi$^*$, and Dongxi Liu$^\S$}%
\vspace{1.6mm}\\
\fontsize{10}{10}\selectfont\itshape
$^\dag$Department of Computer Science, Purdue University, 
305 N. University Street, West Lafayette, IN  47907, USA\\
\fontsize{9}{9}\selectfont\ttfamily\upshape
\{bsamanth, raof, bertino\}@purdue.edu
\vspace{1.6mm}\\
\fontsize{10}{10}\selectfont\normalfont\itshape
$^*$School of Computer Science and Information Technology, RMIT University, Melbourne, Victoria, Australia\\
\fontsize{9}{9}\selectfont\ttfamily\upshape
xun.yi@rmit.edu.au
\vspace{1.6mm}\\
\fontsize{10}{10}\selectfont\normalfont\itshape
$^\S$CSIRO Computational Informatics, Marsfield NSW 2122, Australia\\
\fontsize{9}{9}\selectfont\ttfamily\upshape
dongxi.liu@csiro.au
}

\maketitle
\begin{abstract}
Many techniques for privacy-preserving data mining (PPDM)  
have been investigated over the past decade. Often, the entities involved in 
the data mining process are end-users or organizations with limited computing and storage resources. As 
a result, such entities may want to refrain from participating in the PPDM process.  
To overcome this issue and to take many other benefits of cloud computing, 
outsourcing PPDM tasks to the cloud environment has recently gained 
special attention. We consider the scenario where $n$ entities outsource their 
databases (in encrypted format) to the cloud and ask the cloud to perform the clustering task on their 
combined data in a privacy-preserving manner. We term such a process as privacy-preserving and 
outsourced distributed clustering (PPODC). In this paper, we propose a novel and efficient 
solution to the PPODC problem based on $k$-means clustering algorithm. The main novelty 
of our solution lies in avoiding the secure division operations required in computing 
cluster centers altogether through an 
efficient transformation technique. Our solution 
builds the clusters securely in an iterative fashion and returns the  
final cluster centers to all entities when a pre-determined termination condition holds. 
The proposed solution protects data confidentiality of all the participating 
entities under the standard semi-honest model. 
To the best of our knowledge, ours is the first work to discuss and propose a comprehensive 
solution to the PPODC problem that incurs negligible cost on the participating entities. We theoretically 
estimate both the computation and communication costs of  
the proposed protocol and 
also demonstrate its practical value through experiments on a real dataset. 
\end{abstract}




\section{Introduction} \label{sec:intro}
Clustering is one of the commonly used tasks in various data mining applications. Briefly, 
clustering \cite{data-cluster1,data-cluster2,data-cluster3} is the 
unsupervised classification of data items (or feature vectors) into groups (or clusters) such that similar data items 
reside in the same group. It has 
immense importance in various fields, including information retrieval \cite{ir-cluster}, machine learning \cite{ml-cluster}, 
pattern recognition \cite{pr-cluster}, image analysis \cite{ia-cluster}, and text mining \cite{tm-cluster}. 
Some real-life applications related to clustering include 
categorizing results returned by a search engine in response to
a user's query, grouping persons into categories based on their DNA information, etc. 

In general, if the data involved in clustering belongs to a single entity  
(hereafter referred to as a user), then it can be done in a trivial fashion. 
However, in some cases, multiple users, such as companies, governmental 
agencies, and health care organizations, each holding a dataset, may want to collaboratively perform 
clustering task on their combined data and share the clustering results. Due to 
privacy concerns, users may not be willing to share their data with the other users and thus the 
distributed clustering task\footnote{Note that, a direct application of clustering algorithm locally by
each party is of no use since global evolution of clusters \cite{pp-kmeans1} should be taken
into account.} should be done in a privacy-preserving manner.  This problem, referred to as 
privacy-preserving distributed clustering (PPDC), can be best explained 
by the following example: 
\begin{itemize} 
\item 
Consider two health agencies (e.g., the U.S. CDC and the public health agency of Canada) each holding a dataset containing 
the disease patterns and clinical outcomes of their patients. 
Since both the agencies have their own data collecting methods, 
suppose that they want to cluster their combined datasets and identify interesting clusters that would 
enable directions for better disease control mechanisms. However, 
due to government regulations and the sensitive nature of the data, 
they may not be willing to share their data with one another. Therefore, 
they have to collaboratively perform the clustering task on their joint datasets in a 
privacy-preserving manner. Once the clustering process is done, they can exchange 
necessary information (after proper sanitization) if needed. 
\end{itemize}
The existing PPDC methods (e.g., \cite{ppdc-agg-cluster,pp-kmeans1,pp-kmeans2,bunn-2007}) 
incur significant cost (computation, communication and storage) on the participating users and thus 
they are not suitable if the users do not have sufficient resources to perform the clustering task. This problem  
becomes even more serious when dealing with big data. 
To address these issues, it is more attractive 
for the users to outsource their data as well as the clustering task to the cloud. 
However, the cloud cannot be fully trusted by the users in protecting their data.  
Thus, to ensure data confidentiality, users can encrypt their databases locally (using 
a common public key) and then outsource them to the cloud. Then, the goal is for the cloud 
to perform clustering over the aggregated encrypted data. We refer to the above process 
as \emph{privacy-preserving and outsourced distributed clustering (PPODC)}. 

It is worth noting that if all the encrypted data resides on a single cloud, 
then the only way through which the cloud can perform the clustering task (assuming that users 
do not participate in the clustering process), without ever decrypting 
the data, is when 
the data is encrypted using fully homomorphic encryption schemes (e.g., \cite{gentry-09}). However, recent 
results \cite{gentry-2011} show that fully homomorphic encryption schemes are very expensive and their 
usage in practical applications are decades away. Hence, we believe 
that at least two cloud service providers are required to solve the PPODC problem. 

In this paper, we propose a new and efficient solution to the PPODC problem based on 
the standard $k$-means clustering algorithm \cite{lloyd-1982,fukunaga-1990} by utilizing 
two cloud service providers (say Amazon and Google) which together form a federated 
cloud environment. Our proposed solution protects 
data confidentiality of all the participating users at all times. 
We emphasize that the concept of federated clouds is becoming increasingly popular and is also identified  
as one of the ten High Priority Requirements
for U.S. cloud adoption in the NIST U.S. Government Cloud Computing Technology Roadmap \cite{nist-federated-cloud}. 
Therefore, we believe that developing privacy-preserving solutions under federated 
cloud environment will become increasingly important in the near future.

\subsection{System Model and Problem Definition}\label{sec:prob-def} 
In our problem setting, we consider $n$ users denoted by $U_1,\ldots, U_n$. 
Suppose user $U_i$ holds a database $T_i$ with $m_i$ data records and $l$ 
attributes, for $1\le i \le n$. Consider a scenario where the $n$ users want 
to outsource their databases as well as the $k$-means clustering process on their 
combined databases to a cloud environment. 
In our system model, we 
consider two different entities: (i) the users and (ii) the cloud service providers. 
We assume that the users choose two 
cloud service providers $C_1$ and $C_2$ (say Amazon and Google) to perform 
the clustering task on their combined data. 

In this paper, we explicitly assume 
that $C_1$ and $C_2$ are semi-honest\cite{goldreich-book-gcp} and they do not collude. After proper 
service level agreements with the users, $C_2$ generates a public-secret key pair $(pk, sk)$ 
based on the Paillier cryptosystem \cite{paillier-99} and broadcasts $pk$ to all users 
and $C_1$. A more robust setting would be for $C_1$ and $C_2$ to jointly generate the public key $pk$ based on the 
threshold Paillier cryptosystem (e.g., \cite{damgard-tpc-2001,hazay-2012}) such that the corresponding secret 
key $sk$ is obliviously split between the two clouds. Under this case, the secret key $sk$ is unknown to both clouds 
and only (random) shares of it are revealed to $C_1$ and $C_2$. 
For simplicity, we consider the former asymmetric setting where $C_2$ generates $(pk, sk)$ in the rest of this paper. 
However, our proposed protocol can be easily extended to the above threshold setting without affecting the underlying 
privacy guarantees.

\begin{figure}[!t]
\centering
\epsfig{file=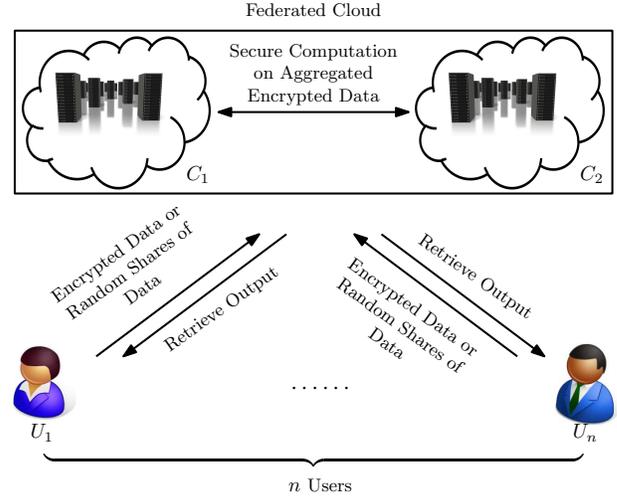, width= .46\textwidth}
\caption{The Proposed PPODC Architecture}
\label{fig:ppodc-arch}
\end{figure}

Given the above system architecture, we assume that user $U_i$ encrypts $T_i$ attribute-wise using $pk$ and outsources  
the encrypted database to $C_1$. Another way to outsource the 
data is that users can split each attribute value in their database into two random shares 
and outsource the shares separately to each cloud (see 
Section \ref{sec:proposed-ppodc} for more details). A detailed 
information flow between different entities in our system model is shown in Figure \ref{fig:ppodc-arch}. 
Having outsourced the data, the main goal of a PPODC protocol is to enable  
$C_1$ and $C_2$ to perform $k$-means clustering over the combined encrypted 
databases in a privacy-preserving manner. More formally, 
we can define a PPODC protocol as follows:
\begin{equation}\label{eq:problem-definition}
\textrm{PPODC}(\langle T_1, \ldots, T_n \rangle, \beta)  \rightarrow (S_1, \ldots, S_n)
\end{equation}
where $\beta$ is a pre-defined threshold value 
agreed upon by all parties. Since $k$-means is an iterative method, we use the
value of $\beta$ to check whether the termination condition holds in each iteration. A more detailed
explanation about the usage of $\beta$ is given in Sections \ref{sec:prelims} and \ref{sec:proposed-section}.
$S_i$ denotes the output received by user $U_i$. Depending on the users' requirements, 
$S_i$ can be the the global cluster centers and/or the final cluster IDs corresponding to the data records of $U_i$. In this paper, 
we consider the former case under which $S_i$'s are the same for all users (however, our protocol can be easily modified 
to handle the latter case). In general, a PPODC protocol should meet the following requirements:  
\begin{itemize}
\item \textbf{Data Confidentiality:} The contents of $U_i$'s database $T_i$ should never 
be revealed to other users, $C_1$ and $C_2$. 
\item \textbf{Accuracy:} The output received by each party (i.e., $S_i$'s ) should be the same as in 
the standard $k$-means algorithm. 
\item \textbf{No Participation of Users:} Since the very purpose of outsourcing is to 
shift the users' load towards the cloud environment, a desirable requirement for any outsourced task is that the  
computations should be totally performed in the cloud. In particular to PPODC, the total clustering process 
should be done by the cloud service providers. 
This will enable the users who do not have enough resources to participate in the clustering task to still get 
the desired results without compromising privacy. 
\end{itemize}
\setlength{\textfloatsep}{0pt}
In certain cases, the user's data (encrypted using his/her own secret key) may have already been stored 
in a cloud (either $C_1$ or different cloud) and he/she want to 
use this data, along with the data from other users, in the clustering task. In the case of the data being stored 
on a different cloud (say $C_3$), the user has to first download and decrypt the data and re-encrypt it under $pk$ and send 
the resulting database  to $C_1$. This might incur heavy cost on the user side, 
especially if the data is large. However, we can address this issue 
using the proxy re-encryption techniques (e.g., \cite{ivan-2003,ateniese-2006}) as follows. 
(i) $C_3$ can directly send the encrypted data of the user to $C_1$, (ii) the user sends a proxy-re-encryption key 
corresponding to his/her secret key and $pk$ to $C_1$, and (iii) $C_1$  
transforms the encrypted data under the user's public key domain into the domain of $pk$ without ever decrypting it using the 
proxy re-encryption key. For ease of presentation, we do not consider the above case 
in the rest of the paper. Instead, we simply assume that all users 
hold their respective databases which 
they can encrypt under 
$pk$ and outsource them to $C_1$.

\subsection{Main Contributions} 
The problem of privacy-preserving clustering over encrypted data in an outsourced environment was addressed 
only recently \cite{liu-2014}. However, the existing method is proposed under a single user setting. To the best of 
our knowledge, there is no existing work that addresses the PPODC problem (i.e., under the multi-user setting). In this paper, 
we propose an efficient and novel PPODC protocol that can enable a group of users to outsource 
their encrypted data as well as the $k$-means clustering task completely to a federated cloud environment 
and ours is the first work along this direction. The main contributions of this work 
are four-fold:
\begin{itemize}
\item We propose new transformations and develop an order-preserving Euclidean distance 
function that enables the proposed PPODC protocol to 
securely assign the data records to the closest clusters, a crucial step in 
each iteration of the $k$-means clustering algorithm. Also, we propose a novel 
transformation for the termination condition that 
enables the PPODC protocol to securely evaluate the termination condition over encrypted data.
\item The proposed solution satisfies all the desirable properties of PPODC mentioned in the previous sub-section. That is, 
it protects the confidentiality of each user's data at all times and outputs the correct result. Also, once 
the user's data is outsourced to the cloud, the user does not need to participate in any computations of the clustering task. 
\item We show that the proposed protocol is secure 
under the standard semi-honest model\cite{goldreich-book-gcp}. 
Also, we theoretically analyze the complexities of the proposed protocol. 
\item 
We demonstrate the practical applicability of our solution through 
extensive experiments using a real-world dataset. 
\end{itemize}
The remainder of this paper is organized as follows. Section \ref{sec:related} discusses the existing related work. 
Section \ref{sec:prelims} presents some definitions and properties related 
to $k$-means clustering algorithm and the Paillier cryptosystem as a background. 
Section \ref{sec:transformation} presents our new transformation techniques. 
Section \ref{sec:proposed-section} discusses our proposed PPODC solution in detail. 
Also, within this section, we analyze the security guarantees and complexities of our solution. 
Section \ref{sec:exp} presents our experimental results on a real-world dataset under different parameter settings.  
Finally, we conclude the paper along with the scope for future research in Section \ref{sec:future}.

\section{Related Work}\label{sec:related}
\subsection{Privacy-Preserving Data Mining (PPDM)}
Our work is closely related 
to the field of privacy-preserving data mining (PPDM) \cite{ppdm1,ppdm2}. 
Several techniques have been proposed for the clustering task under the PPDM model 
(e.g., \cite{ppdc-agg-cluster,pp-kmeans1,pp-kmeans2,bunn-2007}).  
However, we stress that our problem setting is somewhat different  
from the PPDM model. On one hand, under PPDM, each user owns 
a piece of dataset (typically a vertically or horizontally partitioned dataset) 
and the goal is for them to collaboratively perform the clustering task on the combined 
data in a privacy-preserving manner. On the other hand, our work is motivated by the cloud computing model 
where users can outsource their encrypted databases to a federated cloud environment. Under 
our problem setting, the federated cloud performs the clustering 
task over encrypted data and the users do not participate 
in any of the underlying computations. As a result, existing PPDM techniques for the clustering task 
are not applicable to the PPODC problem.

Only recently, researchers have started to focus on the clustering task  
in an outsourced environment (e.g., \cite{upmanyu-2010,liu-2014}). 
The work by Liu et al. \cite{liu-2014} is perhaps the most recent work along this direction. However, their 
solution has the following limitations: (i) it assumes that there is only a single user who wants to perform the 
clustering task on his/her own data and (ii) the user is required to execute 
certain intermediate computations and thus he/she needs to be part of the clustering process. Unlike 
the work in \cite{liu-2014}, our solution is proposed under the multi-user setting and the users can completely outsource 
the computations of the clustering task to a federated cloud environment in a privacy-preserving manner.  

\subsection{Fully Homomorphic Encryption (FHE)}
A straightforward way to solve the PPODC problem is for the users to encrypt 
their data using a fully homomorphic encryption (FHE) scheme, e.g., \cite{gentry-09},  
and outsource the encrypted data to a cloud. Here the secret key should be known only to the users 
(or shared among them). Since FHE allows one to perform arbitrary 
computations over encrypted data without decrypting the data, the cloud can 
perform the clustering task over encrypted data 
and return the encrypted clustering results to the users who can decrypt them. 
Though the FHE schemes enable arbitrary searches or 
operations over encrypted data, such techniques are very expensive and their usage in 
practical applications is decades away. For example, it was shown in \cite{gentry-2011} 
that even for weak security parameters one ``bootstrapping'' operation of a 
homomorphic operation would take at least 30 seconds on a high 
performance machine.

\section{Preliminaries}\label{sec:prelims}
In this section, we first introduce definitions related to cluster centers 
and computation of Euclidean distance between a data record and given cluster. Then, we briefly discuss 
the steps involved in the traditional $k$-means clustering algorithm. Finally, we review 
upon the properties of the threshold Paillier cryptosystem that is adopted in this paper. 

\subsection{Cluster Center}
\begin{definition}\label{def:cluster-center-1}
Let $c = \{t_{1},  \ldots, t_{h}\}$ be a cluster
where $t_{1}, \ldots, t_{h}$ are data records with $l$ attributes. 
Then, the center of cluster $c$ is defined as a vector $\mu_c$ given by \cite{bunn-2007}:
\begin{equation}\label{eq:cluster-center}
\mu_c[s] = \frac{t_{1}[s] + ~\cdots~ + t_{h}[s]}{|c|} = \frac{\lambda_c[s]}{|c|},~\textrm{for}~1 \le s \le l
\end{equation}
where $t_{i}[s]$ denotes the $s^{th}$ attribute value of $t_i$ and $\lambda_c[s]$ denotes the sum 
of $s^{th}$ attribute values of all the data records in cluster $c$, for $1 \le i \le h$. Also,  
$|c|$ denotes the number of data records in $c$. 
\end{definition}
In the above definition, the $s^{th}$ attribute value in $\mu_c$ is equivalent to
the mean of the $s^{th}$ attribute values of all the data records in cluster $c$. Note that, 
if the cluster contains a single data record, then the cluster center is the same as 
the corresponding data record.
\begin{example} Let $c$ be a cluster with three data records \{$t_{1}, t_{2}, t_{3}$\}. Without loss 
of generality, suppose the data records are given as below (assuming $l=5$):
\begin{center}
\begin{tabular}{c l}
$t_{1}$ & = \{0, 2, 1, 0, 3\}\\
$t_{2}$ & = \{1, 1, 3, 4, 2\}\\
$t_{3}$ & = \{0, 1, 0, 2, 0\}\\
\end{tabular}
\end{center}
Then, the center of cluster $c$, based on Definition \ref{def:cluster-center-1},
is given by $\mu_c[1]$ = 0.333, $\mu_c[2]$ = 1.333, $\mu_c[3]$ = 1.333, $\mu_c[4]$ = 2, $\mu_c[5]$ = 1.666.
\hfill $\Box$
\end{example}

\subsection{Computation of Euclidean Distance between $t_i$ and $c$}
We now discuss how to compute the similarity score between a given data record $t_i$ and a cluster $c$. In general, 
the similarity score between any two objects can be computed using one of the standard similarity metrics,
 such as Euclidean distance, Cosine similarity, and Jaccard coefficient. In this
paper, we use the Euclidean distance as the underlying similarity metric since the standard $k$-means 
algorithm is based on this metric \cite{bunn-2007,liu-2014}. 
\begin{definition}\label{def:euclidean-distance}
For any given data record $t_i$ and cluster $c$, let $\mu_c$ denote the cluster center of $c$ (as per 
Definition \ref{def:cluster-center-1}). 
Then the Euclidean distance between $t_i$ and $c$ is given as
\begin{center}
$\|t_i - c\| = \sqrt{\sum\limits_{s=1}^{l} \left ( t_{i}[s] - \mu_c[s] \right )^2}  = \sqrt{\sum\limits_{s=1}^{l} \left ( t_{i}[s] - \frac{\lambda_c[s]}{|c|}\right )^2}$
\end{center}
\end{definition}
\begin{example} Suppose $t_i$ and $\mu_{c}$ are as given below. 
\begin{center}
\begin{tabular}{c l}
$t_i$ & = \{0, 1, 1, 3, 2\}\\
$\mu_{c}$ & = \{0.333, 1.333, 1.333, 2, 1.666\}\\
\end{tabular}
\end{center}
Then, the Euclidean distance between $t_i$ and $c$, based on Definition \ref{def:euclidean-distance},
is $\|t_i - c\| = 1.201$.
\hfill $\Box$
\end{example}
In a similar manner, the Euclidean distance between any two given clusters $c$ and 
$c'$ can be computed using their respective cluster centers. More specifically, $\|c - c'\|$ is 
given as $$\sqrt{\sum_{s=1}^{l} \left ( \mu_c[s] - \mu_{c'}[s] \right )^2 } = \sqrt{\sum_{s=1}^{l} \left ( \frac{\lambda_c[s]}{|c|} - \frac{\lambda_{c'}[s]}{|c'|} \right )^2 }$$
where $\mu_c$ and $\mu_{c'}$ denote the cluster centers of $c$ and $c'$, respectively. Also, 
$|c|$ and $|c'|$ denote the number of data records in $c$ and $c'$, respectively. 

\subsection{Single Party $k$-Means Clustering}\label{sec:kmeans-traditional}
Consider a user $U$ who wants to apply the $k$-means clustering 
algorithm \cite{lloyd-1982,fukunaga-1990} on his/her own database of $m$ records, denoted by 
$\{t_1,\ldots, t_m\}$. Here we assume that $U$ wants to compute $k$ cluster centers, 
denoted by $\mu_{c'_1}, \ldots, \mu_{c'_k}$, as the output. However, other desired values, such as the 
final cluster IDs assigned to each data record can also be part of the output. 
Since $k$-means clustering is an iterative algorithm, $U$ has to input a threshold value to decide 
when to stop the algorithm (termination condition). Without loss of generality, let $\beta$ denote the threshold 
value. Throughout this paper, we assume that the initial set of $k$ clusters are chosen at random (referred 
to as the Initialization step). Note that 
other techniques exist for choosing the initial clusters \cite{bunn-2007}. However, since the goal of this paper is not to investigate 
which initialization technique is better, we simply assume that they are selected at random. 
\begin{algorithm}[t]
\begin{algorithmic}[1]
\REQUIRE User $U$ with $m$ data records $\{t_1,\ldots, t_m\}$ and $\beta$\\
{\textbf{Initialization}}: Select $k$ data records at random and assign them as initial clusters $c_{1},\ldots, c_{k}$ with 
        respective cluster centers as~$\mu_{c_1}, \ldots, \mu_{c_k}$
\FOR{$j=1$ \TO $k$}
  \STATE $c'_{j} \gets \emptyset$
  \STATE $\mu_{c'_j} \gets \{\}$
  \STATE $sum \gets 0$
\ENDFOR

\COMMENT{\textbf{Assignment Stage}}
\FOR{$i=1$ \TO $m$}
     \FOR{$j=1$ \TO $k$}     
       \STATE Compute $\| t_i - c_j\|$       
     \ENDFOR
     \STATE Add $t_i$ to cluster $c'_h$ such that $\|t_i -c_h \|$ is the minimum, for $1 \le h \le k$
\ENDFOR

\COMMENT{\textbf{Update Stage}}
\FOR{$j=1$ \TO $k$}
   \STATE Compute cluster center for $c'_j$ and assign it to $\mu_{c'_j}$
\ENDFOR

\COMMENT{\textbf{Termination Stage} - Compare the old clusters ($c_j$'s) with new clusters ($c'_j$'s) and check whether 
they are close enough}
\STATE $sum \gets \sum\limits_{j=1}^k \|c_j - c'_j\|^2$
\IF{$sum \le \beta$} 
   \STATE Return $\{\mu_{c'_1}, \ldots, \mu_{c'_k}\}$
\ELSE 
    \FOR{$j=1$ \TO $k$}
        \STATE $c_j \gets c'_j$
        \STATE $\mu_{c_j} \gets \mu_{c'_j}$
    \ENDFOR
    \STATE Go to Step 6
\ENDIF
\end{algorithmic}
\caption{\textrm{$k$-means}($\{t_{1}, \ldots, t_{m}\}, \beta) \rightarrow \{\mu_{c'_1}, \ldots, \mu_{c'_k}\}$}
\label{alg:standard-kmeans}
\end{algorithm}
\setlength{\textfloatsep}{0pt}

The main steps involved in the traditional (single party) $k$-means clustering task \cite{lloyd-1982,fukunaga-1990}, using 
the Euclidean distance as the similarity metric, are 
given in Algorithm \ref{alg:standard-kmeans}. Apart 
from the initialization step, the algorithm involves three main stages: (i) Assignment (ii) Update and (ii) Termination. First of 
all, during the initialization step, $k$ data records are selected at random and assigned as
the initial clusters $c_{1},\ldots,c_{k}$ with their centers (or mean 
vectors) denoted by $\mu_{c_1}, \ldots, \mu_{c_k}$, respectively. In the
assignment stage, for each data record $t_i$, the algorithm computes the Euclidean distance between $t_i$ and each cluster $c_j$, 
for $1 \le j \le k$. Then, the algorithm  
identifies the cluster corresponding to the minimum distance as the closest cluster to $t_i$ (say $c_h$) and 
assigns $t_i$ to a new cluster $c'_h$, where $h\in[1, k]$. In the update stage, the algorithm computes the centers
of the new clusters, denoted by $\mu_{c'_1}, \ldots, \mu_{c'_k}$. Finally, in the 
termination stage, the algorithm verifies whether a pre-defined termination 
condition holds. More specifically, the algorithm checks whether the sum of the squared Euclidean 
distances between the current and newly computed clusters  
is less than or equal to the threshold value $\beta$. If the termination
condition holds, then the algorithm halts and returns the new cluster centers as the final 
output. Otherwise, the algorithm continues to the next iteration with the new clusters as input.

\subsection{The Paillier Cryptosystem}\label{sec:tpc}
In this paper, we assume that 
the second cloud service provider $C_2$ generates a public-secret key 
pair $(pk, sk)$ based on the widely used Paillier cryptosystem \cite{paillier-99} which consists of an 
additively homomorphic and probabilistic encryption scheme. Without 
loss of generality, let $E_{pk}(\cdot)$ and $D_{sk}(\cdot)$ denote the encryption and 
decryption functions under Paillier cryptosystem and 
$N$ denote the RSA modulus (or a part of the public key $pk$). 
We emphasize that the Paillier cryptosystem 
exhibits the following properties\cite{paillier-99}:
\begin{itemize}
\item For any $a, b \in~\mathbb{Z}_N$, the encryption scheme is additively homomorphic: $E_{pk}(a)*E_{pk}(b) \bmod N^2 = E_{pk}(a + b \bmod N)$. Due 
to this addition property, the encryption scheme also satisfies the multiplication property  
$E_{pk}(a)^u \bmod N^2 = E_{pk}(a*u \bmod N)$, where $u\in~\mathbb{Z}_N$. 
\item The encryption scheme is semantically secure \cite{goldwasser-1989}. That is, 
given a set of ciphertexts, a computationally bounded adversary cannot deduce any information regarding the 
corresponding plaintexts in polynomial time. 
\end{itemize}
For ease of presentation, we omit the term $\bmod~N^2$ from homomorphic operations in 
the rest of the paper. Also, as mentioned in Section \ref{sec:prob-def}, our proposed protocol  
can be easily extended to the threshold Paillier setting \cite{damgard-tpc-2001} under which $sk$ is obliviously 
generated and shared between $C_1$ and $C_2$ \cite{hazay-2012}.
\section{The Proposed Transformations}\label{sec:transformation}
It is important to note that cluster centers (denoted by $\mu_c$ for a cluster $c$) are represented as vectors and the 
entries in the vectors can be fractional values. Since the encryption schemes typically 
support integer values, we should somehow transform 
the entries of the cluster centers into integer values without affecting their utility in the $k$-means clustering process. Along 
this direction, we first define scaling factors for clusters and then discuss a novel  
order-preserving Euclidean distance function operating over integers. Also, we discuss how to transform 
the termination condition in the $k$-means clustering algorithm with fractional values into an integer-valued one.

\begin{definition}\label{def:center-scaling}
Consider the cluster $c_i$ whose center is denoted by $\mu_{c_i}$ (based on Definition \ref{def:cluster-center-1}). We know 
that $\mu_{c_i}$ is a vector and each entry can be a fractional value with denominator $|c_i|$, for $1 \le i \le k$. We define the scaling 
factor for a cluster $c_i$, denoted by $\alpha_i$, as below:
\begin{equation}\label{eq:scaling-factor}
\alpha_i = \frac{\prod\limits_{j=1}^k |c_j|}{|c_i|} = \prod_{j=1 \wedge j\neq i}^k |c_j|
\end{equation}
Also, we define $\alpha = \prod\limits_{j=1}^k |c_j|$ as the global scaling factor. 
\end{definition}

\subsection{Order-Preserving Euclidean Distance (OPED)}
In the assignment stage of $k$-means clustering, the first step 
is to compute the Euclidean distance between a data record $t_i$ and each cluster $c_j$, denoted 
by $\| t_i - c_j\| =  \sqrt{\sum\limits_{s=1}^{l} \left ( t_{i}[s] - \frac{\lambda_{c_j}[s]}{|c_j|}\right )^2}$. It is clear 
that $\| t_i - c_j\|$ involves fractional value $\frac{\lambda_{c_j}[s]}{|c_j|}$. In order to compute the encrypted value of 
$\|t_i - c_j \|$, we need to avoid such fractional values without affecting the relative ordering  
among the $k$ Euclidean distances $\|t_i - c_1 \|, \ldots, \|t_i - c_k\|$, where $c_1,\ldots, c_k$ denote 
$k$ clusters. Note that since $t_i$ has to be assigned 
to the nearest cluster, it is important to preserve the relative ordering among the computed $k$ Euclidean distances. For this purpose, we 
propose a novel order-preserving Euclidean distance function which works on only integer values. 

We define the order-preserving Euclidean distance (OPED) function between 
a data record $t_i$ and a cluster $c_j$ as follows:
\begin{equation}\label{eq:oped}
\textrm{OPED}(t_i, c_j) = \sqrt{\sum_{s=1}^{l} \left ( \alpha*t_{i}[s] - \alpha_j*\lambda_{c_{j}}[s] \right )^2}
\end{equation}
where $\alpha$ and $\alpha_j$ denote the global and $c_j$'s scaling factors, respectively. Observe  
that all the terms in the above equation are integer values. 
Moreover, following from Definition \ref{def:center-scaling}, we can 
rewrite the above equation as:
\begin{align*}
\textrm{OPED}(t_i, c_j) &= \sqrt{\sum_{s=1}^{l} \left ( \alpha*t_{i}[s] - \frac{\alpha}{|c_j|}*\lambda_{c_j}[s]\right )^2}\\
                       &= \sqrt{\alpha^2* \left ( \sum_{s=1}^{l} \left ( t_{i}[s] - \frac{\lambda_{c_j}[s]}{|c_j|}\right )^2 \right )}\\
                       &= \alpha*\sqrt{\sum_{s=1}^{l} \left ( t_{i}[s] - \frac{\lambda_{c_j}[s]}{|c_j|}\right )^2}\\
                       &= \alpha*\|t_i - c_j\|
\end{align*}
Since $\alpha$ remains constant for any given set of $k$ clusters (in a particular iteration), we claim that 
the above OPED function preserves the relative ordering among cluster centers for any given data record. 
More specifically, given a data record $t_i$ and two clusters $c_j$ and $c_{j'}$, if $\|t_i - c_j\| \ge \|t_i - c_{j'}\|$, 
then it is guaranteed that $\textrm{OPED}(t_i, c_j) \ge \textrm{OPED}(t_i, c_{j'})$, for $1 \le j,j' \le k$ and $ j \neq j'$. 

\subsection{Transformation of the Termination Condition}\label{sec:new-term}
In the $k$-means clustering process (see Algorithm \ref{alg:standard-kmeans}), 
the termination condition is given by: 
\begin{equation}\label{eq:term-cond}
\sum\limits_{j=1}^k \|c_j - c'_j \|^2 \le \beta 
\end{equation}
where ${c_1, \ldots, c_k}$ and ${c'_1, \ldots, c'_k}$ denote the current and 
new set of clusters in an iteration, respectively. Remember that 
$\|c_j - c'_j\| = \sqrt{\sum\limits_{s=1}^l \left ( \frac{\lambda_{c_j}[s]}{|c_j|} - \frac{\lambda_{c'_j}[s]}{|c'_j|} \right )^2}$ 
and clearly it consists of fractional values. In 
order to evaluate this condition over encryption, we first need to 
transform the above termination condition so that all the components are integers. To achieve this, 
we use the following approach. We define a constant scaling factor (denoted by $f$) for the termination condition  in such a way 
that by multiplying Equation \ref{eq:term-cond} with $f^2$, we can cancel all the denominator values. 
More specifically, we define the scaling factor for the termination condition as $f = \prod\limits_{j=1}^k |c_j|*|c'_j|$. Also, 
we define the scaling factor for the cluster pair $(c_j, c'_j)$ as $f_j = \frac{f}{|c_j|*|c'_j|} = \prod\limits_{i=1 \wedge i \neq j}^k |c_i|*|c'_i|$. 
Then we define the new termination condition as follows:
\begin{equation}\label{eq:new-term-cond}
\sum_{j=1}^k \sum_{s=1}^l \left ( |c'_j|*f_j*\lambda_{c_j}[s] - |c_j|*f_j*\lambda_{c'_j}[s] \right )^2 \le f^2*\beta
\end{equation}
Observe that the above equation consists of only integer values. Now we need to show that evaluating 
the above equation is the same as evaluating Equation \ref{eq:term-cond}. First, we divide 
the above equation by $f^2$ on both sides of the inequality. Note that since $f^2$ remains constant in 
a given iteration, multiplication of Equation \ref{eq:new-term-cond} by $f^2$ has no 
effect on the inequality. Precisely, Equation \ref{eq:new-term-cond} can be rewritten as: 
$$\sum_{j=1}^k \sum_{s=1}^l \frac{\left ( |c'_j|*f_j*\lambda_{c_j}[s] - |c_j|*f_j*\lambda_{c'_j}[s] \right )^2}{f^2} \le \beta $$
Given this, the left-hand side of the above equation can be expanded as follows: 
\begin{align*}
\sum_{j=1}^k \sum_{s=1}^l \left ( \frac{|c'_j|*f_j*\lambda_{c_j}[s]}{f} - \frac{|c_j|*f_j*\lambda_{c'_j}[s]}{f} \right )^2 &\\
= \sum_{j=1}^k \sum_{s=1}^l \left ( \frac{|c'_j|*\lambda_{c_j}[s]}{|c_j|*|c'_j|} - \frac{|c_j|*\lambda_{c'_j}[s]}{|c_j|*|c'_j|}\right )^2 \\
= \sum_{j=1}^k \sum_{s=1}^l \left ( \frac{\lambda_{c_j}[s]}{|c_j|} - \frac{\lambda_{c'_j}[s]}{|c'_j|} \right )^2 \\
= \sum_{j=1}^k \|c_j - c'_j\|^2
\end{align*}
Based on the above discussions, it is clear that evaluating the inequality $\sum_{j=1}^k \|c_j - c'_j \|^2 \le \beta$ is the same 
as evaluating Equation \ref{eq:new-term-cond}. Hence, in 
our proposed PPODC protocol, we consider Equation \ref{eq:new-term-cond} as the termination condition of $k$-means 
clustering and evaluate it in a privacy-preserving manner.

\section{The Proposed Solution}\label{sec:proposed-section}
In this section, we first discuss a set of privacy-preserving 
primitives. Then, we present our novel PPODC protocol that utilizes the 
above transformation techniques and the privacy-preserving 
primitives as building blocks. 

As mentioned in Section \ref{sec:prob-def}, in this paper we consider 
two semi-honest and non-colluding cloud service 
providers $C_1$ and $C_2$ under the Paillier cryptosystem \cite{paillier-99}. 
More specifically, 
$C_2$ generates a pair of public-secret key pair $(pk, sk)$ based on the Paillier's scheme such 
that $sk$ is kept private whereas 
the corresponding public key $pk$ is broadcasted. 
\subsection{Privacy-Preserving Primitives}
We discuss a set of privacy-preserving primitives under 
the above two-party (i.e., $C_1$ and $C_2$) computation model.
\begin{itemize}
\item Secure Multiplication (SMP): Given that $C_1$ holds $\langle E_{pk}(a), E_{pk}(b) \rangle$ and $C_2$ holds $sk$, 
where $\langle a, b\rangle$ is unknown to both $C_1$ and $C_2$, the 
goal of the SMP protocol is to compute $E_{pk}(a*b)$. During the execution of SMP, no information regarding 
the contents of $a$ and $b$ is revealed to $C_1$ and $C_2$. 
\item Secure Squared Euclidean Distance (SSED): In this protocol, $C_1$ holds two 
encrypted vectors $E_{pk}(X) = \langle E_{pk}(x[1], \ldots, E_{pk}(x[l])\rangle$ 
and $E_{pk}(Y) = \langle E_{pk}(y[1]), \ldots, E_{pk}(y[l]) \rangle$. 
The goal of SSED is to compute the encryption of the squared Euclidean distance between $X$ and $Y$. 
Specifically, the output is $E_{pk}((\| X - Y\|)^2)$. The SSED protocol should  
reveal neither the contents of $X$ and $Y$ nor the Euclidean distance between them to $C_1$ and $C_2$. 
\item Secure Squared Order-Preserving Euclidean Distance (SSED$_\textrm{OP}$): Given that 
$C_1$ holds an encrypted data record, denoted by $E_{pk}(t_i)$, and an encrypted cluster, denoted by $E_{pk}(c_h)$, the goal 
of the SSED$_\textrm{OP}$ protocol is for $C_1$ and $C_2$ to jointly compute $E_{pk}((\textrm{OPED}(t_i, c_h))^2)$.  Here 
$E_{pk}(c_h) =  \langle E_{pk}(\lambda_{c_h}), E_{pk}(|c_h|)\rangle$ and 
$E_{pk}(\lambda_{c_h}) = \langle E_{pk}(\lambda_{c_h}[1]), \ldots, E_{pk}(\lambda_{c_h}[l])\rangle$. Note that OPED$(t_i, c_h)$ denotes 
the Euclidean distance between $t_i$ and cluster $c_h$ based on the order-preserving Euclidean distance function defined  
in Equation \ref{eq:oped}.  
At the end of this protocol, the output $E_{pk}((\textrm{OPED}(t_i, c_h))^2)$ 
is revealed only to $C_1$ and no other information is revealed to $C_1$ and $C_2$. 
\item Secure Least Significant Bit (SLSB): Given that $C_1$ holds $E_{pk}(z)$, where $z$ is unknown 
to both parties, the goal of SLSB is to compute encryption of the least 
significant bit (LSB) of $z$. The output $E_{pk}([z]_1)$ is revealed only to $C_1$, 
where $[z]_1$ denotes the LSB of $z$. During the execution of the SLSB protocol, 
no contents regarding $z$ is revealed to $C_1$ and $C_2$.
\item Secure Comparison (SC): Given that $C_1$ holds $\langle E_{pk}(a), E_{pk}(b)\rangle$, the 
goal of SC is to securely compare $a$ and $b$. 
The output of SC is $E_{pk}(\gamma)$, where $\gamma = 1$ 
if $a \le b$, and 0 otherwise. At the end, $E_{pk}(\gamma)$ is known only to $C_1$ and no 
other information is revealed to $C_1$ and $C_2$. 
\item Secure Minimum (SMIN): Assume that $C_1$ holds  
$\langle E_{pk}(a), E_{pk}(s_a)\rangle$ and $\langle E_{pk}(b), E_{pk}(s_b)\rangle$, where $s_a$ 
and $s_b$ are the secrets associated with integers $a$ and $b$, respectively. The goal 
of SMIN is to compute the encryption of minimum value between $a$ and $b$, denoted 
by $E_{pk}(\min(a,b))$. In addition, it computes the encryption of the secret  
corresponding to the minimum value. More specifically, the final output of SMIN  
is $(T, I)$,  and it will be revealed only 
to $C_1$.  Here $T =E_{pk}(\min(a, b))$, and $I = E_{pk}(s_a)$ if $a$ is the minimum value, and $I = E_{pk}(S_b)$ otherwise. During 
SMIN, no information regarding $a$ and $b$ is revealed to $C_1$ and $C_2$. 
\item Secure Minimum out of $k$ Numbers (SMIN$_k$): 
In this protocol, we assume 
that $C_1$ holds $k$ encrypted integers and $C_2$ holds $sk$. The goal of SMIN$_k$ is to securely 
identify the location corresponding to the minimum value among the $k$ numbers. More specifically, if 
$j^{th}$ integer is the minimum number among the $k$ values, then the output of SMIN$_k$ is 
an encrypted vector such that its $j^{th}$ component is $E_{pk}(1)$ and the rest are encryptions of 0, 
where $ j \in [1, k]$. The SMIN$_k$ protocol should not reveal any information 
regarding the contents of $k$ numbers (e.g., the minimum value or 
the location corresponding to it, etc.) to $C_1$ and $C_2$. 
The SMIN$_k$ protocol can be treated as a generalization of SMIN in which the secrets associated 
with the $k$ integers represent their locations. 
\end{itemize}
Several solutions have been proposed for most of the above 
privacy-preserving primitives. Recently, Yousef et al. \cite{yousef-icde14} discussed 
efficient implementations for SMP and SSED.  Also, an efficient 
solution to SLSB was proposed in \cite{bksam-asiaccs13}. In the 
rest of this paper,  SMP and SSED refer to the implementations given 
in \cite{yousef-icde14}. Similarly, by SLSB, we 
refer to the implementation given in \cite{bksam-asiaccs13}. We now propose   
efficient implementations to SSED$_\textrm{OP}$, SC, SMIN, 
and SMIN$_k$. 

\subsubsection{The SSED$_\textrm{OP}$ Protocol}
We discuss a novel solution to the SSED$_\textrm{OP}$ problem using 
the SMP and SSED protocols as sub-routines. The main steps 
involved in the proposed SSED$_\textrm{OP}$ protocol are highlighted in Algorithm \ref{alg:ssed-op}. 
We assume that $C_1$ holds $\langle E_{pk}(c_1), \ldots, E_{pk}(c_k)\rangle$ and $C_2$ holds $sk$, 
where $c_1, \ldots, c_k$ denote $k$ clusters and $E_{pk}(c_h) = \langle E_{pk}(\lambda_{c_h}), E_{pk}(|c_h|)\rangle$. Note 
that $E_{pk}(\lambda_{c_h}) = \langle E_{pk}(\lambda_{c_h}[1]), \ldots, E_{pk}(\lambda_{c_h}[l])\rangle$. The goal 
of SSED$_\textrm{OP}$ is to securely compute $E_{pk}((\textrm{OPED}(t_i, c_h))^2)$ for a given input 
$E_{pk}(t_i)$ and $E_{pk}(c_h)$, where $1 \le h \le k$.

\begin{algorithm}[!t]
\begin{algorithmic}[1]
\REQUIRE $C_1$ has $E_{pk}(t_i)$, ${E_{pk}(c_h) = \langle E_{pk}(\lambda_{c_h}), E_{pk}(|c_h|)\rangle}$
\STATE $C_1$ and $C_2$:
\begin{enumerate}     
      \item[(a).] $b_h \gets \textrm{SMP}_{k-1}(\tau_h)$, where $\tau_h = \cup_{j=1 \wedge j \neq h}^k E_{pk}(|c_j|)$                               
     \item[(b).] $b' \gets \textrm{SMP}(b_h, E_{pk}(|c_h|))$           
     \item[(c).] \textbf{for} $1 \le s \le l$ \textbf{do:}
          \begin{itemize}
                  \item $a_i[s] \gets \textrm{SMP}(b', E_{pk}(t_i[s]))$
                  \item $a'_h[s] \gets \textrm{SMP}(b_h, E_{pk}(\lambda_{c_h}[s]))$
          \end{itemize}  
    \item[(d).] $E_{pk}((\textrm{OPED}(t_i, c_h))^2) \gets \textrm{SSED}(a_i, a'_h)$
\end{enumerate}
\end{algorithmic}
\caption{SSED$_\textrm{OP}(E_{pk}(t_i), E_{pk}(c_h))$}
\label{alg:ssed-op}
\end{algorithm}

To start with, $C_1$ and $C_2$ securely compute the scaling factor for cluster 
$c_h$ (in encrypted format based on Equation \ref{eq:scaling-factor}) 
using the extended secure multiplication protocol, denoted by SMP$_{k-1}$, that takes $k-1$ encrypted inputs and 
multiplies them (within encryption). 
Specifically, they jointly compute $b_h = \textrm{SMP}_{k-1}(\tau_h)$, where 
$\tau_h = \cup_{j=1 \wedge j \neq h}^k E_{pk}(|c_j|)$. The important observation here is
that $b_h = E_{pk}(\prod_{j=1 \wedge j \neq h}^k |c_j|) = E_{pk}(\alpha_h)$, 
where $\alpha_h$ is the scaling factor for cluster $c_h$ as defined in Equation 3. Then $C_1$ and $C_2$ 
securely multiply $b_h$ with $E_{pk}(|c_h|)$ using SMP to get $b' = \textrm{SMP}(b_h, E_{pk}(c_h)) = 
E_{pk}(|c_1|*\ldots*|c_k|) = E_{pk}(\alpha)$, 
where $\alpha$ is the global scaling factor. 
After this, for $1 \le s \le l$, $C_1$ and $C_2$ jointly compute two encrypted vectors as follows: 
\begin{align*}
a_i[s] &= \textrm{SMP}(b', E_{pk}(t_i[s])) = E_{pk}(\alpha * t_i[s])\\
a'_h[s] &= \textrm{SMP}(b_h, E_{pk}(\lambda_{c_h}[s])) = E_{pk}(\alpha_h * \lambda_{c_h}[s])
\end{align*}
Finally, with the two encrypted vectors $a_i$ and $a'_h$ as $C_1$'s input, $C_1$ 
and $C_2$ jointly compute the encrypted squared Euclidean distance between them using the SSED protocol. 
More specifically, the output of SSED$(a_i, a'_h)$ is $E_{pk}(\sum_{s=1}^l (\alpha * t_i[s] - \alpha_h * \lambda_{c_h}[s])^2)$. 
Following from Equation 4, it is clear that the output SSED$(a_i, a'_h)$
is equivalent to $E_{pk}((\textrm{OPED}(t_i, c_h))^2)$.

\subsubsection{The Secure Comparison (SC) Protocol}
Given that $C_1$ holds $\langle E_{pk}(a), E_{pk}(b) \rangle$ and $C_2$ holds $sk$, 
the goal of SC is to return $E_{pk}(\gamma)$ such that $\gamma=1$ iff $a \le b$, 
and $\gamma=0$ otherwise. During SC, neither the contents of $(a,b)$ nor the comparison 
result $\gamma$ should be disclosed to $C_1$ and $C_2$.

\begin{table*}[!ht]
\centering
\renewcommand{\arraystretch}{1.4}
\setlength{\tabcolsep}{15pt}
\caption{Truth Table for $a \leq b$} \label{tab:truthTable}
\begin{tabular}{ccc|c}
\hline
\hline
$w = (a < N/2)$ & $x = (b < N/2)$ & $y = (b - a \bmod{N} < N/2)$ & $\gamma = (a \leq b)$\tabularnewline
\hline
\hline
0 & 1 & * & 0\tabularnewline
\hline
1 & 0 & * & 1\tabularnewline
\hline
0 & 0 & 0 & 0\tabularnewline
\hline
0 & 0 & 1 & 1\tabularnewline
\hline
1 & 1 & 0 & 0\tabularnewline
\hline
1 & 1 & 1 & 1\tabularnewline
\hline
\end{tabular}
\end{table*}

We emphasize that it is desirable to have an SC protocol 
whose efficiency does not rely on the bit length of the input integers (i.e., $a$ and $b$) to be compared. We 
now discuss about such a solution constructed by combining  
SLSB \cite{bksam-asiaccs13} with the ideas proposed by Nishide et al. \cite{nishide-2007}. The SC 
solution proposed in \cite{nishide-2007} is based on the secret sharing scheme\cite{Shamir79}. However, it 
is also applicable to our problem domain upon simple modifications. 

In what follows, we briefly describe how $C_1$ and $C_2$ can securely compute 
the encryption of $\gamma$, given $\langle E_{pk}(a), E_{pk}(b) \rangle$ as $C_1$'s private 
input, using the ideas proposed 
in \cite{nishide-2007}. According to \cite{nishide-2007}, the 
value of comparison result $\gamma$ solely depends on the following $3$ predicates: 
$w: a < N/2$, $x: b < N/2$, and $y: b - a \bmod{N} < N/2$. More specifically, $\gamma$ is given 
as: 
\begin{align}\label{eq:sc-formula}
\gamma &= w\overline{x} \vee \overline{w} \mbox{ } \overline{x}y \vee wxy \nonumber\\
&= w(1-x) \vee (1-w)(1-x)y \vee wxy \nonumber \\
&= -x(w + y - 2wy) + (w + y - wy)
\end{align}
More specifically, 
all possible combinations of $(w, x, y)$ and their corresponding $\gamma$ values are given in Table \ref{tab:truthTable}, 
where $*$ denotes either bit 0 or 1.
The main challenge here is that how $C_1$ can compute $E_{pk}(w)$, $E_{pk}(x)$ and $E_{pk}(y)$ given $E_{pk}(a)$ 
and $E_{pk}(b)$. 

As highlighted in \cite{nishide-2007}, one can notice that 
$a \in \{0, 1, \cdots, \frac{(N-1)}{2}\}$ iff $[2a \bmod{N}]_{1} = 0$. Similarly, 
$a \in \{\frac{(N-1)}{2} + 1, \cdots, N-1\}$ iff $[2a \bmod{N}]_{1} = 1$, 
where $[2a]_{1}$ signifies the least significant bit (LSB) of $2a \bmod N$. 
That is, $w = 1$ (implying that $a < N/2$) if and only if the LSB of $(2a \bmod{N})$ is 0, i.e., 
$w \Leftrightarrow 1 - [2a \bmod{N}]_{1}$. Similar conclusions can be drawn for $x$ and $y$. 
Consider the case of computing $E_{pk}(w)$ from $E_{pk}(a)$. First, $C_1$ can locally 
compute $E_{pk}(2a)$. Then, in order 
to compute the encrypted LSB of $2a$, $C_1$ and $C_2$ jointly involve 
in the SLSB protocol\cite{bksam-asiaccs13}.  At the end of this step, only $C_1$ knows 
$E_{pk}([2a]_1)$. Now $C_1$ can locally compute $E_{pk}(w) = E_{pk}(1)*E_{pk}([2a]_1)^{N-1} = E_{pk}(1  - [2a]_1)$. 
In a similar fashion, $C_1$ can compute $E_{pk}(x)$ and $E_{pk}(y)$. Finally, 
$C_1$ (with the help of $C_2$) can compute $E_{pk}(\gamma)$ based 
on Equation \ref{eq:sc-formula}. Note that this step explicitly requires SMP as a building block.

\subsubsection{The SMIN Protocol} 
Let $a$ and $b$ be two integers in $\mathbb{Z}_{N}$, and 
$s_{a}$ and $s_{b}$ be their associated secrets, respectively. For example, if $a$ and 
$b$ correspond to two data records, then their secrets can correspond to the record identifiers. 
Suppose that $\min(a,b)$ denote the minimum value between $a$ and $b$ and that 
$s_{\min(a,b)}$ denote the secret corresponding to $\min(a,b)$. 
Given that $\langle E_{pk}(a), E_{pk}(s_{a})\rangle$ and $\langle E_{pk}(b), E_{pk}(s_{b})\rangle$ as $C_1$'s private 
input, the goal of SMIN is to securely compute $\langle E_{pk}(\min(a, b)), E_{pk}(s_{\min(a,b)})\rangle$ as 
the final output and it should be known only to $C_1$.

We now discuss a simple to SMIN based on the SC protocol. As discussed above, 
at the end of SC protocol, $C_1$ knows $E_{pk}(\gamma)$, where $\gamma$ denotes  
the comparison result of functionality $a \le b$. 
Given $E_{pk}(\gamma)$, $C_1$ can securely compute the encryption of the 
minimum value between $a$ and $b$, 
i.e., $E_{pk}(\min(a,b))$, using the 
following formulation: 
$$\min(a,b) = \gamma*a + (1 - \gamma)*b$$ 
More specifically, using the secure multiplication (SMP) protocol, $C_1$ with 
input $\langle E_{pk}(a), E_{pk}(b), E_{pk}(\gamma) \rangle$ and $C_2$ with $sk$ can compute 
$E_{pk}(\gamma*a)$ and $E_{pk}(\gamma*b)$. Note that the output of SMP will be known 
only to $C_1$. After this, $C_1$ can compute $E_{pk}(\min(a,b))$  
as $E_{pk}(\gamma*a)*E_{pk}(b)* E_{pk}(\gamma*b)^{N-1}$ locally. 

In a similar manner, apart from the encrypted 
minimum value, $C_1$ and $C_2$ compute the encryption of the secret 
associated with the minimum value. More specifically,  
they compute $E_{pk}(s_{\min(a,b)})$ using the following formulation:
$$s_{\min(a,b)} = \gamma*s_a + (1 - \gamma)*s_b$$

\begin{example} 
Suppose $C_1$ holds $E_{pk}(7)$ and $E_{pk}(4)$ (i.e., $a=7$ and $b=4$). Without 
loss of generality, let $E_{pk}(s_1)$ and $E_{pk}(s_2)$ denote their respective secrets. 
It is clear that the SC protocol returns $E_{pk}(0)$ (i.e., $\gamma=0$) 
as output to $C_1$ since $a \le b$ does not hold in this example. 
The output of SMIN is $E_{pk}(\min(7, 4)) = E_{pk}(\gamma*a + (1 - \gamma)*b) = E_{pk}(b) = E_{pk}(4)$ 
and $E_{pk}(s_{\min(a,b)}) = E_{pk}(s_2)$.
\hfill $\Box$
\end{example}

\subsubsection{The SMIN$_k$ Protocol}
Given $k$ encrypted integers, the SMIN$_k$ protocol  
computes an encrypted vector $\Gamma$ of length $k$ such that the entry 
corresponding to the minimum value is an encryption of 1 and the rest are encryptions of 0. We now 
discuss a novel SMIN$_k$ protocol constructed using the SMIN protocol as a building block. The overall 
steps in the proposed SMIN$_k$ protocol are give in Algorithm \ref{alg:smink}.

Suppose $E_{pk}(d_1), \ldots, E_{pk}(d_k)$ denote the list of $k$ encrypted integers and $i$ denotes  
the index (or location) of integer $d_i$ in the list, for $1 \le i \le k$. Initially, using the SMIN protocol, $C_1$ with 
input $(E_{pk}(d_1), E_{pk}(1))$ and $(E_{pk}(d_2), E_{pk}(2))$ and $C_2$ can compute $T = E_{pk}(\min(d_1, d_2))$ and 
$I = E_{pk}(s_{\min(d_1,d_2)})$, where $s_{\min(d_1,d_2)}$ denotes the location of the minimum value between 
$d_1$ and $d_2$. Note that the output of the SMIN protocol is known only to $C_1$. After 
this $C_1$ with input $(T, I)$ and $(E_{pk}(d_3), E_{pk}(3))$ can 
compute $E_{pk}(\min(d_1, d_2, d_3))$ and $E_{pk}(s_{\min(d_1,d_2,d_3)})$ using SMIN. The 
above process is repeated until $I = E_{pk}(s_{\min(d_1, \ldots, d_k)})$ is computed (known only to $C_1$), 
where $s_{\min(d_1, \ldots, d_k)}$ denotes the index (or location) corresponding to the minimum 
value among the $k$ input values. This process is shown as Step 1 in Algorithm \ref{alg:smink}.

After this, $C_1$ and $C_2$ perform the following set of operations:
\begin{itemize}
\item 
$C_1$ computes $E_{pk}(i - s_{\min(d_1, \ldots, d_k)})$ and randomizes it to get $\phi[i]  = E_{pk}(r_i*(i - s_{\min(d_1, \ldots, d_k)}))$, where 
$r_i$ denotes a random number in $\mathbb{Z}_N$ and $1 \le i \le k$. Observe that exactly 
one of the entries in $\phi$ is equal to encryption of 0 (i.e., when $i = s_{\min(d_1, \ldots, d_k)}$) and 
the rest are encryptions of random values. Hereafter, we use the notation $r \in_R \mathbb{Z}_N$ 
to denote  a random number $r$ in $\mathbb{Z}_N$. 
\item $C_1$ computes $u = \pi(\phi)$ and sends it to 
$C_2$. Here $\pi$ is a random permutation function known only to 
$C_1$. 
\item Upon receiving $u$, $C_2$ decrypts it component-wise using $sk$ to get  
$u'[i] = D_{sk}(u[i])$. After this, $C_2$ generates an encrypted vector $U$ as follows. 
If $u'[i] = 0$, then $U[i] = E_{pk}(1)$, and $E_{pk}(0)$ otherwise. $C_2$ sends $U$ to $C_1$. 
\item Finally, $C_1$ gets the desired encrypted vector $\Gamma$ as output by performing an inverse permutation 
on $U$.
\end{itemize}

\begin{algorithm}[!t]
\begin{algorithmic}[1]
\REQUIRE $C_1$ holds $(E_{pk}(d_1), \ldots, E_{pk}(d_k))$ and $\pi$; $C_2$ holds $sk$.
\STATE $C_{1}$ and $C_{2}$:
		\begin{enumerate}
                 \item[(a).] $(T, I) \leftarrow \textrm{SMIN}((E_{pk}(d_{1}), E_{pk}(1)), (E_{pk}(d_{2}), E_{pk}(2)))$
	         \item[(b).] {\bf for} $i = 2$ to $k - 1$ {\bf do:}
		        \begin{itemize}             
                             \item $(T, I) \leftarrow \textrm{SMIN}((T, I), (E_{pk}(d_{s+1}), E_{pk}(s+1)))$
	                \end{itemize}
		\end{enumerate}
	
\STATE $C_{1}$:
		\begin{enumerate}
		\item[(a).] $\Delta \leftarrow I^{N-1}$
		\item[(b).] {\bf for} $i = 1$ to $k$ {\bf do:}
			\begin{itemize}
			\item $\Delta'[i] \leftarrow E_{pk}(i) *\Delta$ \label{step:GammaI}
			\item $\phi[i] \leftarrow \Delta'[i]^{r_{i}}$, where $r_{i} \in_R \mathbb{Z}_{N}$ \label{step:GammaIPrime}
			\end{itemize}

		\item[(c).] $u \leftarrow \pi(\phi)$; send $\phi$ to $C_{2}$
		\end{enumerate}
		
\STATE $C_{2}$:
		\begin{enumerate}
		\item[(a).] Receive $u$ from $C_{1}$	
		\item[(b).] {\bf for} $i = 1$ to $k$ {\bf do:}
			\begin{itemize}
			\item $u'[i] \leftarrow D_{sk}(u[i])$
			\end{itemize}
		\item[(c).] {\bf for} $i = 1$ to $k$ {\bf do:}
			\begin{itemize}
			\item {\bf if} $u'[i] = 0$ {\bf then} $U[i] \leftarrow E_{pk}(1)$
			\item {\bf else} $U[i] \leftarrow E_{pk}(0)$
			\end{itemize}

		\item[(d).] Send $U$ to $C_{1}$
		\end{enumerate}
		
	\STATE $C_{1}$:
		\begin{enumerate}
		\item[(a).] Receive $U$ from $C_{2}$
		\item[(b).] $\Gamma \leftarrow \pi^{-1}(U)$
		\end{enumerate}
	\end{algorithmic}

\caption{SMIN$_k(E_{pk}(d_1), \ldots, E_{pk}(d_k)) \rightarrow \Gamma$}
\label{alg:smink}	
\end{algorithm}

\begin{example}
Let $k=5$ and the input to SMIN$_k$ be $\langle E_{pk}(3), E_{pk}(6), E_{pk}(13), E_{pk}(2), E_{pk}(9) \rangle$. The output 
at the end of Step 1 in the proposed SMIN$_k$ protocol is $\langle T, I \rangle = \langle E_{pk}(2), E_{pk}(4) \rangle$ and it will be known only to $C_1$. 
Note that since `2' is the minimum among the five input values, the output 
of Step 1 is encryption of `2' and encryption of the location corresponding to `2'  in the input list (i.e., $s_{\min(3, 6, 13, 2, 9)} = 4$). 
After this, $C_1$ computes $\phi[1] = E_{pk}(r_1*(1-4))$, $\phi[2] = E_{pk}(r_2*(2-4))$, $\phi[3] = E_{pk}(r_3*(3-4))$, 
$\phi[4] = E_{pk}(r_4*(4-4))$, and $\phi[5] = E_{pk}(r_5*(5-4))$. Without loss 
of generality, let the random permutation function $\pi$ (known only to $C_1$) be as follows. 
\begin{table}[!h]
\centering
\renewcommand{\arraystretch}{1}
\begin{tabular}{l l l l l l l l}
$i$ &=& \quad 1 & \quad 2 & \quad 3 & \quad 4 & \quad 5 \\
& & \quad $\downarrow$ & \quad $\downarrow$ & \quad $\downarrow$ & \quad $\downarrow$ & \quad $\downarrow$ \\
$\pi(i)$~~&=& \quad 2 & \quad 5 & \quad 1 & \quad 3 & \quad 4\\
\end{tabular}
\end{table} 
Now $C_1$ computes $u = \pi(\phi) = \langle \phi[3], \phi[1], \phi[4], \phi[5], \phi[2] \rangle $ and 
sends the resulting vector $u$ to $C_2$. Upon receiving, $C_2$ 
decrypts it using $sk$ and identifies that $D_{sk}(u[3]) = 0$. Note 
that the rest of the values are random numbers. Then $C_2$ computes $U = \langle E_{pk}(0), E_{pk}(0), E_{pk}(1), E_{pk}(0), E_{pk}(0) \rangle$ 
and sends it to $C_1$. Finally, $C_1$ computes the final output as  
$\Gamma = \pi^{-1}(U) = \langle E_{pk}(0), E_{pk}(0), E_{pk}(0), E_{pk}(1), E_{pk}(0) \rangle$.
\hfill $\Box$
\end{example}

\subsection{The Proposed PPODC Protocol}\label{sec:proposed-ppodc}
In this sub-section, we discuss our proposed PPODC protocol which is based on the 
standard $k$-means algorithm discussed in Section \ref{sec:kmeans-traditional}. 
As mentioned in Section \ref{sec:prob-def}, 
our system model consists of $n$ users denoted by $U_1, \ldots, U_n$. User $U_j$ holds 
a database $T_j$ of $m_j$ data records with $l$ attributes, for $1 \le j \le n$. Without loss 
of generality, 
let the aggregated database be $T = \bigcup_{j=1}^n T_j = \{t_1,\ldots, t_m\}$,  
where $m = \sum_{j=1}^n m_j$ denotes the total number of records in $T$. 
For simplicity, let $t_1\ldots t_{m_1}$ belong to $U_1$, $t_{m_1 + 1}, \ldots, t_{m_1 + m_2}$ 
belong to $U_2$, and so on.  
We assume that all users agree upon using two cloud service 
providers $C_1$ and $C_2$ for outsourcing their respective databases as well as the 
$k$-means clustering task. Remember that, in our system model, $C_2$ generates a public-secret key pair $(pk, sk)$ 
based on the Paillier cryptosystem \cite{paillier-99} and the public key $pk$ 
is sent to all users and $C_1$.

After the users outsource their data (encrypted under $pk$) to $C_1$, the goal of PPODC is to enable   
$C_1$ and $C_2$ to jointly compute the global cluster centers using the aggregated encrypted data in 
a privacy-preserving manner. At a high level, our protocol computes 
the global cluster centers in an iterative manner until 
the pre-defined termination condition (given in Equation \ref{eq:new-term-cond}) holds. 

The overall steps involved in the proposed PPODC protocol are given in
Algorithms \ref{alg:main-protocol} and \ref{alg:setc}. The main steps 
are shown in Algorithm \ref{alg:main-protocol}. 
Briefly, the PPODC protocol consists of the following three stages:
\begin{itemize}
\item \textbf{Stage 1 - Secure Data Outsourcing:}\\ 
During this stage, each user $U_j$ has to securely outsource an encrypted version 
of his/her database $T_j$ to $C_1$. To minimize the data 
encryption costs of users, we achieve data outsourcing 
through randomization techniques. Note that this stage is run only once.  
At the end of this stage, only $C_1$ knows the (attribute-wise) encryptions 
of the $n$ databases. 
\item \textbf{Stage 2 - Secure Computation of New Clusters:}\\ 
In this stage, $C_1$ initially selects $k$ data records at random (from the aggregated 
encrypted records) 
and assigns them as initial clusters (this step is the same as the initialization step 
in the traditional $k$-means algorithm). Then, $C_1$ and $C_2$ jointly assign each data record to a new cluster. 
After this, they compute the new cluster centers in encrypted format. The main goal of this stage is similar 
to the assignment and update stages given in Algorithm 1. 
\item \textbf{Stage 3 - Secure Termination or Update:}\\
Upon computing the new cluster centers (in encrypted format), $C_1$ and $C_2$ 
securely verify whether the sum of the squared Euclidean distances between 
the current and new clusters is less than or equal to $\beta$ (termination condition based 
on Equation \ref{eq:new-term-cond}). Here $\beta$ denotes 
the pre-defined threshold value agreed upon by all the participating users. If the termination 
condition holds, then the protocol terminates returning the new cluster 
centers as the final output. Otherwise, $C_1$ and $C_2$ update 
the current clusters to the new clusters and repeat Stages 2 and 3. 
\end{itemize}
We emphasize that Stage 1 of PPODC is executed only once whereas Stages 2 and 3 
are run in an iterative manner. We now discuss the steps in each of these three stages in detail. 
\subsubsection{Stage 1 - Secure Data Outsourcing (SDO)} 
Data are typically encrypted before being outsourced for privacy reasons. However, to avoid computation overhead 
on the users side due to having to encrypt their data, we consider 
the following approach for data outsourcing. User $U_j$ generates two random shares for 
each attribute value of his/her data record $t_{i}$. Precisely, for the $s^{th}$ attribute of data 
record $t_{i}$, $U_j$ generates two random shares $(t_{i}^1[s], t_{i}^2[s])$ given by $t_{i}^1[s] = t_{i}[s] + r_{i}[s] \bmod N$ and 
$t_{i}^2[s] =  N - r_{i}[s]$, where $r_{i}[s] \in_R \mathbb{Z}_N$ and 
$1 \le s \le l$. Observe that $t_{i}[s] = t_{i}^1[s] + t_{i}^2[s] \bmod N$. $U_j$ outsources 
the random shares $t_i^1[s]$ and $t_i^2[s]$ to $C_1$ and $C_2$, respectively, instead of encrypting the database attribute-wise 
and outsourcing it to 
$C_1$.  Thus, we are able to avoid 
heavy encryption costs on the users during the data outsourcing step. Here 
we assume that there exist secure communication channels, which can be established using standard mechanisms such as SSL, 
between user $U_j$ and the two clouds 
$C_1$ and $C_2$. Each user $U_j$ sends the random shares of his/her data to $C_1$ and $C_2$ separately through 
the secure communication channels. 

After receiving the random shares for all the data records,  
$C_2$ computes $E_{pk}(t_i^2[s])$ and sends it 
to $C_1$. Then $C_1$ computes $E_{pk}(t_i[s]) = E_{pk}(t_i^1[s])* E_{pk}(t_i^2[s])$, for $1 \le i \le m$ and $1 \le s \le l$.   
 
\begin{algorithm}[!htbp]
\begin{algorithmic}[1]
\REQUIRE $U_j$ holds a private database $T_j$ with $m_j$ data records, $sk$ is known only to $C_2$ \\
\COMMENT{\textbf{Stage 1 - Secure Data Outsourcing}}
\STATE \textbf{for} $1 \le i \le m$\textbf{:}
\begin{enumerate}\itemsep=0pt
       \item[(a).] \textbf{for} $1 \le s \le l$\textbf{:}
             \begin{itemize}\itemsep=0pt
                   \item \textbf{if} $t_i \in T_j$ \textbf{then:}
                        \begin{itemize}
                             \item $U_j$ computes $t_{i}^1[s] = t_{i}[s] + r_{i}[s] \bmod N$,  $t_{i}^2[s] =  N - r_{i}[s]$, and $r_{i}[s]$ is 
a random number in $\mathbb{Z}_N$; sends $t_{i}^1[s]$ to $C_1$ and $t_{i}^2[s]$ to $C_2$
                        \end{itemize}
                   \item $C_2$ sends $E_{pk}(t_{i}^2[s])$ to $C_1$
                   \item $C_1$ computes $E_{pk}(t_{i}[s]) \gets E_{pk}(t_{i}^1[s])* E_{pk}(t_{i}^2[s])$
             \end{itemize}  
\end{enumerate}
\COMMENT{\textbf{Stage 2 - Secure Computation of New Clusters}}
\STATE $C_1$: 
\begin{enumerate}\itemsep=0pt  
       \item[(a).] Select $k$ records at random and assign them to initial clusters denoted 
by $E_{pk}(\lambda_{c_1}), \ldots, E_{pk}(\lambda_{c_k})$, where $c_1, \ldots, c_k$ denote the current clusters 
       \item[(b).] $E_{pk}(|c_h|) \gets E_{pk}(1)$, for $1 \le h \le k$   
\end{enumerate}
\STATE \textbf{for} $1 \le i \le m$ \textbf{do:}
\begin{enumerate}
       \item[(a).] $C_1$ and $C_2$:
            \begin{itemize}\itemsep=0pt                                      
                   \item $E_{pk}(d_{i}[h]) \gets \textrm{SSED}_{\textrm{OP}}(E_{pk}(t_{i}), E_{pk}(c_h))$, for $1 \le h \le k$, where 
$E_{pk}(c_h) = \langle E_{pk}(\lambda_{c_h}), E_{pk}(|c_h|)\rangle$
                   \item $\Gamma_{i} \gets \textrm{SMIN}_k(E_{pk}(d_{i}[1]), \ldots, E_{pk}(d_{i}[k]))$
                   \item $\Lambda_{i,h}[s] \gets \textrm{SMP}(\Gamma_{i,h}, E_{pk}(t_{i}[s]))$, for $1 \le h \le k$ and $1 \le s \le l$
            \end{itemize}       
\end{enumerate}
\STATE $C_1$:
\begin{enumerate}\itemsep=0pt
      \item[(a).] \textbf{for} $1 \le h \le k$ \textbf{do:}
           \begin{itemize}\itemsep=0pt                   
                   \item $ W_h[s] \gets  \prod\limits_{i=1}^m \Lambda_{i,h}[s]$, for $1 \le s \le l$ 
                   \item $E_{pk}(|c'_h|) \gets \prod\limits_{i=1}^m \Gamma_{i,h}$ 
           \end{itemize}
\end{enumerate}
\COMMENT{\textbf{Stage 3 - Secure Termination or Update}}
\STATE $\gamma \gets \textrm{SETC}(\Omega, \Omega')$, where $\gamma$ denotes 
whether the termination condition holds or not, $\Omega = \{\langle E_{pk}(\lambda_{c_1}), E_{pk}(|c_1|) \rangle \ldots, \langle E_{pk}(\lambda_{c_k}), E_{pk}(|c_k|) \rangle\}$ 
and $\Omega' = \{\langle W_1, E_{pk}(|c'_1|) \rangle \ldots, \langle W_k, E_{pk}(|c'_k|) \rangle\}$
\STATE \textbf{if} $\gamma = 1$ \textbf{then}, for $1 \le h \le k$ and $1 \le s \le l$
\begin{enumerate}
         \item[(a).] $C_1$: 
                      \begin{itemize}
                          \item $O_h[s] \gets W_h[s]*E_{pk}(r'_h[s])$ and $\delta_h \gets E_{pk}(|c'_h|)*E_{pk}(r''_h)$, 
where $r'_h[s]$ and $r''_h$ $\in_R~\mathbb{Z}_N$                          
                          \item Send $O_h[s]$ and $\delta_h$ to 
$C_2$; $r'_h[s]$ and $r''_h$ to each user $U_j$
                      \end{itemize}
         \item[(b).] $C_2$: Send $O'_h[s] \gets D_{sk}(O_h[s])$ and $\delta'_h \gets D_{sk}(\delta_h)$ to each user $U_j$ \\                                                  
         \hspace*{-0.6cm}\textbf{else}, for $1 \le h \le s$
                      \begin{itemize}\itemsep=1pt
                          \item $E_{pk}(\lambda_{c_h}) \gets W_h$ and $E_{pk}(|c_h|) \gets E_{pk}(|c'_h|)$  
                          \item Go to Step 3
                      \end{itemize}  
\end{enumerate}
\STATE $U_j$, \textbf{foreach} received pairs $\langle O'_h, r'_h)$ and $\langle \delta'_h, r''_h\rangle$ $\textbf{do}$:
\begin{enumerate}
                         \item[(a).] $\lambda_{c'_h}[s] = O'_h[s] - r'_h[s] \bmod N$, $1 \le s \le l$
                         \item[(b).] $|c'_h| \gets \delta'_h - r''_h \bmod N$ 
                         \item[(c).] $ \mu_{c'_h}[s] \gets \frac{\lambda_{c'_h}[s]}{|c'_h|}$ and $S_j \gets S_j \cup \mu_{c'_h}$  
\end{enumerate}
\end{algorithmic}
\caption{PPODC$(\langle T_1, \ldots, T_n \rangle, \beta)  \rightarrow (S_1, \ldots, S_n)$}
\label{alg:main-protocol}
\end{algorithm}

\subsubsection{Stage 2 - Secure Computation of New Clusters (SCNC)}
Given the (attribute-wise) encrypted versions of users data, during Stage 2, $C_1$ and $C_2$ jointly 
compute the new cluster centers in a privacy-preserving manner. To start with, $C_1$ 
randomly selects $k$ encrypted data records (from the aggregated data) and assigns them as initial clusters. More 
specifically, the $k$ encrypted data records are assigned to $E_{pk}(\lambda_{c_1}),\ldots, E_{pk}(\lambda_{c_k})$, respectively. 
For example, if the 3rd data record is selected as the first cluster $c_1$, then $E_{pk}(\lambda_{c_1}[s])$ 
is set to $E_{pk}(t_3[s])$, for $1 \le s \le l$. Also, $E_{pk}(|c_h|)$ is set to 
$E_{pk}(1)$ since each initial cluster $c_h$ consists of only one data record, for $1 \le h \le k$.   

For each encrypted data record $E_{pk}(t_i)$, $C_1$ and $C_2$ compute the squared 
Euclidean distance between $t_i$ and all the clusters based on the order-preserving Euclidean 
distance function given in Equation \ref{eq:oped}. To achieve this, $C_1$ and $C_2$ jointly 
execute the SSED$_\textrm{OP}$ protocol with $E_{pk}(t_i)$ and $E_{pk}(c_h)$ as $C_1$'s 
private input, for $1 \le i \le m$ and $1 \le h \le k$, where $E_{pk}(c_h) = \langle E_{pk}(\lambda_{c_h}), E_{pk}(|c_h|) \rangle$. 
The output of SSED$_\textrm{OP}$ is denoted by $E_{pk}(d_i[h])$. Note 
that $d_i[h] =  (\textrm{OPED}(t_i, c_h))^2$. Now, $C_1$ and $C_2$ jointly execute 
the following set of operations: 
\begin{itemize}
\item For $1 \le i \le m$, with the $k$ encrypted distances as $C_1$'s private input to 
the secure minimum out of $k$ numbers (SMIN$_k$) protocol, $C_1$ and $C_2$ compute an 
encrypted bit vector $\Gamma_i$. The important observation here is that $\Gamma_{i,g}$ is an encryption of 1 iff  
$d_{i}[g]$ is the minimum distance among $\langle d_i[1], \ldots, d_i[k]\rangle$. In this case, 
$t_i$ is closest to cluster $c_g$, where $1 \le g \le k$. The rest of the values in $\Gamma_i$ are 
encryptions of 0. Note that the output of SMIN$_k$, i.e., $\Gamma_i$, is known only to $C_1$. 
\item After this, $C_1$ and $C_2$ securely multiply $\Gamma_{i,h}$ with $E_{pk}(t_{i}[s])$ using 
the secure multiplication (SMP) sub-protocol. Precisely, $C_1$ and $C_2$ compute 
$\Lambda_{i,h}[s]=$~SMP$(\Gamma_{i,h},  E_{pk}(t_i[s]))$. 
The observation here is that since $\Gamma_{i,g} = E_{pk}(1)$ only if $t_i$ is closest to 
cluster $c_g$, $\Lambda_{i,g} = E_{pk}(t_i)$ denoting that $t_i$ is assigned to new cluster $c'_g$. Also, $\Lambda_{i,h}$ is 
a vector of encryptions of 0, for $1 \le h \le k$ and $h \neq g$.
\end{itemize}
Next, $C_1$ computes the new cluster centers locally by performing homomorphic operations 
on $\Lambda_{i,h}$ and $\Gamma_{i,h}$ as follows:
\begin{itemize}
\item Compute (in encrypted format) the $s^{th}$-component of the numerator for the center of new cluster $c'_h$ as 
$W_h[s] =  \prod\limits_{i=1}^m \Lambda_{i,h}[s]$, for $1 \le h \le k$ and $1 \le s \le l$. The observation here 
is $W_h[s] = E_{pk}(\lambda_{c'_h}[s])$. Remember that $\mu_{c'_h}[s]  = \frac{\lambda_{c'_h}[s]}{|c'_h|}$, 
where $\mu_{c'_h}$ denotes the center of  cluster $c'_h$.
\item Compute the number of data records (in the encrypted format) that belong to the new cluster 
$c'_h$ as $E_{pk}(|c'_h|) = \prod\limits_{i=1}^m \Gamma_{i,h}$, for $1 \le h \le k$.
\end{itemize}

\subsubsection{Stage 3 - Secure Termination or Update (STOU)}
Given the new clusters (in encrypted format) resulting from Stage 2, the goal 
of Stage 3 is for $C_1$ and $C_2$ to verify whether the termination condition (based 
on Equation \ref{eq:new-term-cond}) holds in a privacy-preserving 
manner. If the termination condition holds, the new cluster centers are returned as the final 
output to each user. Otherwise, the entire iterative process (i.e., Stages 2 and 3) is repeated by using the new clusters  
as the current clusters. The current and new clusters are given by 
$\Omega = \{\langle E_{pk}(\lambda_{c_1}), E_{pk}(|c_1|) \rangle, \ldots, \langle E_{pk}(\lambda_{c_k}), E_{pk}(|c_k|) \rangle\}$ 
and ${\Omega' = \{\langle W_1, E_{pk}(|c'_1|) \rangle \ldots, \langle W_k, E_{pk}(|c'_k|) \rangle\})}$, respectively.

First, by using the current and new clusters, $C_1$ and $C_2$ need to securely evaluate 
the termination condition (SETC) based on Equation 6. The main steps involved in SETC are given  
in Algorithm \ref{alg:setc} which we explain in detail below: 
\begin{itemize}
\item $C_1$ and $C_2$ compute $\tau_i = E_{pk}(|c_i|* |c'_i|)$ using 
$\langle E_{pk}(c_i), E_{pk}(c'_i) \rangle$ as $C_1$'s private input to 
the SMP sub-protocol, for $1 \le i \le k$. The output $\tau_i$ is known only to $C_1$. 
\item By using $\tau_i$'s, they compute $V_i = \textrm{SMP}_{k-1}(\tau'_i)$, where 
$\tau'_i = \cup_{j=1 \wedge j \neq i}^k \tau_j$. Here SMP$_{k-1}$ denotes 
the SMP protocol with $k-1$ encrypted inputs that need to 
be securely multiplied. More specifically, $V_i = E_{pk}(\prod_{j=1\wedge j \neq i}^k |c_i|* |c'_i|)$, for 
$1 \le i \le k$. The important observation here 
is $$V_i =  E_{pk}\left ( \prod_{j=1\wedge j \neq i}^k |c_i|* |c'_i| \right ) = E_{pk}(f_i)$$ 
where $f_i$ is the scaling factor for cluster pair $(c_i, c'_i)$ defined in Section \ref{sec:new-term}. 
Then, they compute an encrypted value $Z_i$ as
$$Z_i = \textrm{SMP}(V_i, V_i) = E_{pk}(f_i^2)$$
\item After this, they securely multiply $V_1$ and $\tau_1$ using SMP  
protocol. The output of this step is
$$V = \textrm{SMP}(V_1, \tau_1) = E_{pk}\left ( \prod_{j=1}^k |c_j|* |c'_j| \right ) = E_{pk}(f)$$
where $f$ is the scaling factor for the termination condition as defined in 
Section \ref{sec:new-term}. Then, they compute $$Y = \textrm{SMP}(V, V) = E_{pk}(f^2)$$.
\item For $1 \le i \le k$, $C_1$ and $C_2$ securely 
multiply each component in the current and new clusters with $|c'_i|$ and $|c_i|$, respectively. 
More specifically, for $1 \le i \le k$ and $1 \le s \le l$, they compute 
\begin{align*}
G_i[s] &= \textrm{SMP}(E_{pk}(\lambda_{c_{i}}[s]), E_{pk}(|c'_i|))\\
       &= E_{pk}(\lambda_{c_{i}}[s]*|c'_i|)\\
G'_i[s] &= \textrm{SMP}(W_i[s], E_{pk}(|c_i|))\\
        &= E_{pk}(\lambda_{c'_i}[s] * |c_i|)
\end{align*}
Note that $W_i[s]$ computed in Stage 2 is equivalent to  $E_{pk}(\lambda_{c'_i}[s])$.
\item Now, by using the secure squared Euclidean distance (SSED) protocol 
with input vectors $G_i$ and $G'_i$, $C_1$ and $C_2$ jointly compute 
$H_i = \textrm{SSED}(G_i, G'_i)$. Precisely, they compute the encryption 
of squared Euclidean distance between vectors in $G_i$ and $G'_i$ given by,  
$$H_i = E_{pk}\left ( \sum_{s=1}^l (\lambda_{c_i}[s]*|c'_i| - \lambda_{c'_i}[s] * |c_i|)^2 \right )$$ 
\item Given $Z_i$ and $H_i$, $C_1$ and $C_2$ can securely multiply them 
to get $H'_i = \textrm{SMP}(H_i, Z_i) = E_{pk}\left ( f_i^2 * \sum_{s=1}^l (\lambda_{c_i}[s]*|c'_i| - \lambda_{c'_i}[s] * |c_i|)^2 \right )$
\end{itemize} 
At the end of the above process, $C_1$ has $Y = E_{pk}(f^2)$ and $H'_i$, for $1 \le i \le k$. 
Now $C_1$ locally computes: 
\begin{align*}
R &= Y^{\beta} = E_{pk}(f^2*\beta)~~~\textrm{and}\\       
L &= \prod_{i=1}^k H'_i\\
        &= \prod_{i=1}^k E_{pk}\left ( f_i^2 * \sum_{s=1}^l (\lambda_{c_i}[s]*|c'_i| - \lambda_{c'_i}[s] * |c_i|)^2 \right ) \\
        & = E_{pk}\left ( \sum_{i=1}^k \sum_{s=1}^l (\lambda_{c_i}[s]*f_i*|c'_i| - \lambda_{c'_i}[s] *f_i* |c_i|)^2 \right )\\        
\end{align*}
At this point, $C_1$ has encryptions of the integers corresponding to both the left-hand and right-hand sides of the termination 
condition given in Equation 6. Therefore, the goal is to now securely compare them using 
the secure comparison (SC) protocol.  
More specifically, by using $L$ and $R$ as $C_1$'s private input to the SC protocol, $C_1$ and $C_2$ securely 
evaluate the termination condition: 
$$\sum_{i=1}^k \sum_{s=1}^l \left ( \lambda_{c_i}[s]*f_i*|c'_i| - \lambda_{c'_i}[s] *f_i* |c_i| \right )^2 \le f^2*\beta$$ 
The output is $E_{pk}(\gamma) = \textrm{SC}(L,R)$, where $\gamma=1$ iff the termination condition 
holds, and $\gamma=0$ otherwise. Note that $E_{pk}(\gamma)$ is known only to $C_1$. After this, $C_1$ sends 
$E_{pk}(\gamma)$ to $C_2$, who decrypts it and forwards the value of $\gamma$ 
to $C_1$. 
\begin{algorithm}[!t]
\begin{algorithmic}[1]
\REQUIRE $C_1$ has $\Omega = \{\langle E_{pk}(\lambda_{c_1}), E_{pk}(|c_1|) \rangle, \ldots, \langle E_{pk}(\lambda_{c_k}),$\linebreak$E_{pk}(|c_k|) \rangle\}$, 
${\Omega' = \{\langle W_1, E_{pk}(|c'_1|) \rangle \ldots, \langle W_k, E_{pk}(|c'_k|) \rangle\}}$
\STATE $C_1$ and $C_2$:
\begin{enumerate}
      \item[(a).] $\tau_i \gets \textrm{SMP}(E_{pk}(|c_i|), E_{pk}(|c'_i|))$, for $1 \le i \le k$
      \item[(b).] \textbf{for} $1 \le i \le k$ \textbf{do:}
           \begin{itemize}
                   \item $V_i \gets \textrm{SMP}_{k-1}(\tau'_i)$, where $\tau'_i = \cup_{j=1 \wedge j \neq i}^k \tau_j$ 
                   \item $Z_i \gets \textrm{SMP}(V_i, V_i)$  
           \end{itemize}  
     \item[(c).] $V \gets \textrm{SMP}(V_1, \tau_1)$
     \item[(d).] $Y \gets \textrm{SMP}(V, V)$ 
      
     \item[(e).] \textbf{for} $1 \le i \le k$ and $1 \le s \le l$ \textbf{do:}
          \begin{itemize}
                  \item $G_i[s] \gets \textrm{SMP}(E_{pk}(\lambda_{c_i}[s]), E_{pk}(|c'_i|))$
                  \item $G'_i[s] \gets \textrm{SMP}(W_i[s], E_{pk}(|c_i|))$
          \end{itemize}  
    \item[(f).] $H_i \gets \textrm{SSED}(G_i, G'_i)$, for $1 \le i \le k$
    \item[(g).] $H'_i \gets \textrm{SMP}(H_i, Z_i)$, for $1 \le i \le k$    
\end{enumerate}   
\STATE $C_1$: $L \gets \prod\limits_{i=1}^k H'_i$ and $R \gets Y^\beta$
\STATE $C_1$ and $C_2$: 
\begin{enumerate}
      \item[(a).] $E_{pk}(\gamma) \gets \textrm{SC}(L, R)$, note that the output of SC 
is known only to $C_1$
        
\end{enumerate}
\STATE $C_1$: Send $E_{pk}(\gamma)$ to $C_2$
\STATE $C_2$: Decrypt $E_{pk}(\gamma)$ and send $\gamma$ to $C_1$
\end{algorithmic}
\caption{SETC$(\Omega, \Omega')$}
\label{alg:setc}
\end{algorithm}
\begin{table*}[!thbp]
\footnotesize
\centering
\renewcommand{\arraystretch}{1.3}
\setlength{\tabcolsep}{2pt}
\caption{Online and offline computational costs for different stages in \ppodcopt}
\label{tb:computation-analysis}
\begin{tabular}{|l | c| c|}
\hline
\textbf{Stage} & \textbf{Online} & \textbf{Offline}\\
\hline
Stage 1 (one-time) & $6m*l$ mul. & $2m*l$ exp. \\
\hline
Stage 2 (per iteration) & $m*(2l*k + l + 17k -4 \myfloor{k}{2} -14) + k*(l+1) + 1$ exp. & $m*(7l*k + 3l + 32k -7\myfloor{k}{2} -29) + k*(3l+1) + 1$ exp.\\
\hline
Stage 3 (per iteration) & $k*(2k+5l)+9$ exp. & $k*(4k + 9l+ 2) + 20$ exp.\\
\hline
\end{tabular}
\vspace*{-0.4cm}
\end{table*}
\setlength{\textfloatsep}{0pt}

\noindent Finally, once the termination condition has been securely evaluated, $C_1$ locally proceeds as 
follows:
\begin{itemize}
\item If $\gamma=1$ (i.e., when the termination condition holds), the newly computed 
clusters are the final clusters which need to be sent to each user $U_j$. For this purpose, 
$C_1$ takes the help of $C_2$ to obliviously decrypt the results related to the new 
cluster centers. More specifically, $C_1$ initially picks two sets of random numbers $\langle r'_h[s], r''_h \rangle$  and 
computes $O_h[s] = W_h[s]*E_{pk}(r'_h[s]) = E_{pk}(\lambda_{c'_h}[s] + r_h[s] \bmod N)$ and 
$\delta_h = E_{pk}(|c'_h|)* E_{pk}(r''_h) = E_{pk}(|c'_h|+ r''_h \bmod N)$, for $1 \le h \le k$ and $1 \le s \le l$. 
After this, $C_1$ 
sends $O_h[s]$ and $\delta_h$ to $C_2$. In addition, $C_1$ sends 
$r'_h[s]$ and $r''_h$ to each user $U_j$ (through separate and secure communication channels). 
\item For $1 \le s \le l$, $C_2$ successfully decrypts the received encrypted values using his/her secret share $sk$ 
to get $O'_h[s] = D_{sk}(O_h[s])$ and $\delta'_h = D_{sk}(\delta_h)$ 
which it forwards to each user $U_j$ (through separate and secure communication channels). Observe that, due 
to the randomization by $C_1$, the values of $O'_h[s]$ and $\delta'_h$ are random numbers in $\mathbb{Z}_N$ from 
$C_2$'s perspective.
\item Upon receiving the entry pairs $\langle O'_h, r'_h \rangle$ and $\langle \delta'_h, r''_h\rangle$, 
each user $U_j$ removes the random factors to get $\lambda_{c'_h}[s] = O'_h[s] - r'_h[s] \bmod N$ and 
$|c'_h| = \delta'_h - r''_h \bmod N$, for $1 \le h \le k$ and $1 \le s \le l$. Finally, $U_j$ computes 
the final cluster center $\mu_{c'_h}$ as $\mu_{c'_h}[s] = \frac{\lambda_{c'_h}[s]}{|c'_h|}$ and 
adds it to his/her resulting set $S_j$.
\item On the other hand, when $\gamma =0$, then $C_1$ locally updates the current clusters 
to new clusters by setting $E_{pk}(\lambda_{c_h}) = W_h$ and $E_{pk}(|c_h|) = E_{pk}(|c'_h|)$, for $1 \le h \le k$. 
After this, the above process is repeated in an iterative manner until the termination condition holds. 
That is, the protocol goes to Step 3 of Algorithm \ref{alg:main-protocol} and  
executes Steps 3 to 6 with the updated cluster centers as input.  
\end{itemize}

\subsection{Security Analysis of PPODC under the Semi-honest Model}
In this section, we show that the proposed PPODC protocol is secure under 
the standard semi-honest model \cite{goldreich-book-gcp,Goldreichenc}. Informally speaking, we stress 
that all the intermediate values seen by $C_1$ and $C_2$ in PPODC are either encrypted or pseudo-random numbers.

First, in the data outsourcing process (i.e., Step 1 of Algorithm \ref{alg:main-protocol}), the values 
received by $C_1$ and $C_2$ are either random or pseudo-random values in $\mathbb{Z}_N$. At the 
end of the data outsourcing step, only $C_1$ knows the encrypted data records of all users and  
no information regarding the contents of $T_j$ (the database of user $U_j$) is revealed to $C_2$. 
Therefore, as long as the underlying encryption scheme is semantically secure (which is also 
the case in the Paillier cryptosystem \cite{hazay-2012}), the aggregated encrypted databases do 
not reveal any information to $C_1$. Hence, no information is revealed to $C_1$ and $C_2$ during 
Stage 1 of PPODC. 

The implementations of SMP, SSED, and SLSB sub-protocols given in \cite{bksam-ppknn-tech,bksam-asiaccs13} are 
proven to be secure under the semi-honest model \cite{goldreich-book-gcp}. Also, the SC protocol given 
in \cite{nishide-2007} is secure under the semi-honest model. In the proposed 
SSED$_\textrm{OP}$ protocol, the computations are based on using either 
SMP or SSED as a sub-routine. As a result, SSED$_\textrm{OP}$ can be proven to be 
secure under the semi-honest model. Further, since SMIN and SMIN$_k$ are directly 
constructed from SC, the security proofs for them directly follow from the security 
proof of SC given in \cite{nishide-2007}. In summary, the 
privacy-preserving primitives utilized in the proposed 
PPODC protocol are secure under the semi-honest model. 

We emphasize that the computations involved in Stages 2 and 3 of PPODC are performed 
by either $C_1$ locally or using one of the privacy-preserving primitives as a 
sub-routine. In the former case, $C_1$ operates on encrypted data locally. In the latter case, 
the privacy-preserving 
primitives utilized in our protocol are secure under the semi-honest model. Also, 
it is important to note that the output of a privacy-preserving primitive which is fed 
as input to the next primitive is in encrypted format. Since we use a semantically secure  
Paillier encryption scheme \cite{paillier-99}, all the encrypted results (which are revealed only 
to $C_1$) from 
the privacy-preserving primitives do not reveal any information to $C_1$. Note that the secret key $sk$ is 
unknown to $C_1$. 
Hence, by Composition Theorem \cite{Goldreichenc}, 
we claim that the sequential composition of the privacy-preserving primitives lead to  
Stages 2 and 3 in our proposed PPODC protocol and are secure under the semi-honest model.  
Putting everything together, it is clear that PPODC is secure under 
the semi-honest model. 


\subsection{Performance Analysis of PPODC}\label{sec:perf-anal}
First of all, we emphasize that a direct implementation of the 
proposed PPODC protocol is likely to be inefficient. To address this issue, we propose 
two strategies to boost its performance: {\em(i) offline computation} and {\em(ii) reusability of intermediate results}. 
In what follows, we extensively analyze the performance of PPODC based on these two strategies. 

In the Paillier cryptosystem \cite{paillier-99}, encryption of an integer $a\in \mathbb{Z}_N$ is given 
by $E_{pk}(a) = g^a*r^N \bmod N^2$, where $g$ is the generator, $N$ is the RSA modulus, and $r$ is a 
random number in $\mathbb{Z}_N$. It is clear that Paillier's encryption scheme requires two expensive 
exponentiation operations. In this paper, we assume $g = N+1$ (a commonly used setting that provides the same security guarantee as 
the original Paillier cryptosystem) as this allows for a more 
efficient implementation of Paillier encryption\cite{damgard-2010}. More specifically, 
when $g = N+1$, we have 
\begin{eqnarray}\label{eq:paillier-ext}
E_{pk}(a) &=& (N+1)^a*r^N \bmod N^2 \nonumber \\
         &=& (a*N + 1)*r^N \bmod N^2
\end{eqnarray}
As a result, an encryption under Paillier is reduced to one exponentiation operation. Our main observation from 
Equation \ref{eq:paillier-ext} is that the encryption cost under Paillier can be further reduced as follows. 
The exponentiation operation (i.e., $r^N \bmod N^2$) in the 
encryption function can be computed in an offline phase and thus the online cost of computing $E_{pk}(a)$ is reduced 
to two (inexpensive) multiplication operations\footnote{The time that takes to perform one exponentiation  
under $\mathbb{Z}_{N^2}$ is equivalent to $\log_2 N$ multiplication operations. Therefore, exponentiation is considered 
to be an expensive operation in comparison to multiplication.}. Additionally, encryption of random 
numbers, 0s and 1s can be precomputed by the corresponding party (i.e., $C_1$ or $C_2$) 
as they are independent of the underlying protocol. 

We emphasize that the actual online computation costs (with an offline phase) 
of the privacy-preserving primitives used in our protocol  can be much less than their costs without an offline phase. For 
example, consider the secure multiplication (SMP) primitive with $E_{pk}(a)$ and $E_{pk}(b)$ as $C_1$'s private input. 
During the execution of SMP, $C_1$ has to initially randomize the inputs and send them to $C_2$. That is, $C_1$ has to 
compute $E_{pk}(a)*E_{pk}(r_1) = E_{pk}(a + r_1 \bmod N)$ and $E_{pk}(b)*E_{pk}(r_2) = E_{pk}(b + r_2 \bmod N)$, where 
$r_1$ and $r_2$ are random numbers in $\mathbb{Z}_N$. This clearly 
requires $C_1$ to compute two encryptions: $E_{pk}(r_1)$ and $E_{pk}(r_2)$. However, since $r_1$ and $r_2$ are integers 
chosen by $C_1$ at random, the computation of $E_{pk}(r_1)$ and $E_{pk}(r_2)$ is independent 
of any specific instantiation of SMP. That is, $C_1$ can precompute 
$E_{pk}(r_1)$ and $E_{pk}(r_2)$ during the offline phase and thus boosting its online computation time. 
In a similar manner, $C_1$ and $C_2$ can precompute certain intermediate results in each 
privacy-preserving primitive. 

To better understand the performance improvements due to the above offline computation strategy, we have analyzed 
the offline and online computation costs of each privacy-preserving primitive (for a single execution) used in PPODC,  
separately. The results are given in Table \ref{tb:primitives-analysis}. Here $l$ denotes number 
of attributes and $k$ denotes number of desired clusters. From our analyses, following 
from Table \ref{tb:primitives-analysis}, we observed that 
the actual online computation cost (with an offline phase) of each primitive is improved by at least 50\% in comparison 
to its online computation cost without an offline phase. 

\begin{table}[h]
\footnotesize
\centering
\renewcommand{\arraystretch}{1.5}
\setlength{\tabcolsep}{3pt}
\caption{Online and offline computation costs of privacy-preserving primitives (measured 
in terms of number of exponentiations)}
\label{tb:primitives-analysis}
\begin{tabular}{|l | c| c|}
\hline
\textbf{Primitive} & \textbf{Online} & \textbf{Offline}\\
\hline
SMP & 2 & 4\\
\hline
SSED & $3l$ & $4l$ \\
\hline
SSED$_\textrm{OP}$ & $2k + 7l$ & $4k + 12l$\\
\hline
SLSB & 1 & 3\\ 
\hline
SC & 7 & 17 \\
\hline
SMIN & 14 & 30\\
\hline
SMIN$_k$ & $16k - 4 \myfloor{k}{2} -14$ & $31k - 7\myfloor{k}{2} - 29$\\
\hline
\end{tabular}
\end{table}

An important observation in PPODC is that some of the intermediate 
results (apart from those computed during the offline phase) computed in earlier steps  
can be reused in later computations without affecting the security. This leads  
to our second performance improvement strategy - reusability of intermediate results. This would be better 
illustrated by the following example. Consider that $C_1$  with private input 
$\langle E_{pk}(a) , E_{pk}(b_1) \rangle$ and $C_2$ jointly want to compute $E_{pk}(a*b_1)$ using SMP. During this process, 
$C_1$ initially computes $E_{pk}(a + r \bmod N)$ and $E_{pk}(b_1 + r_1 \bmod N)$ and sends them to $C_2$, 
where $r$ and $r_1$ are random numbers in $\mathbb{Z}_N$. Upon receiving the ciphertexts, $C_2$ 
decrypts them to get $a + r \bmod N$ and $b_1 + r_1 \bmod N$ and proceeds with the rest of the computations involved 
in SMP. At a later stage, suppose $C_1$ with private input $\langle E_{pk}(a) , E_{pk}(b_2) \rangle$ 
and $C_2$ want to compute $E_{pk}(a*b_2)$. The key observation here is that 
$C_1$ can compute and send only $E_{pk}(b_2 + r_2 \bmod N)$ to $C_2$, where $r_2$ is a random 
number in $\mathbb{Z}_N$. That is, there is no need for $C_1$ to again 
compute $E_{pk}(a + r \bmod N)$ and send that to $C_2$. 
After receiving  $E_{pk}(b_2 + r_2 \bmod N)$ from $C_1$, $C_2$ can decrypt it 
to get $b_2 + r_2 \bmod N$ and use the intermediate result $a + r \bmod N$ already computed 
in the previous step to proceed with further computations of SMP. The above example clearly demonstrates that 
reusability of intermediate results can save both computation and communication costs.  

\begin{table*}[t]
\footnotesize
\centering
\renewcommand{\arraystretch}{1.5}
\setlength{\tabcolsep}{2pt}
\caption{Communication costs of \ppodcopt}
\label{tb:comm-analysis}
\begin{tabular}{|l | c|}
\hline
\textbf{Stage} & \textbf{Communication Cost (in bits)}\\
\hline
Stage 1 (one-time) & $4m*l*K$\\
\hline
Stage 2 (per iteration) & $(4m*l*k + 2l*k + 2*m*l + 21m*k - 2m*\myfloor{k}{2} - 19m + 1)*2K$\\
\hline
Stage 3 (per iteration) & $(k*(3k + 6l + 2) + 15)*2K$\\
\hline
\end{tabular}
\end{table*}

By taking both the above two strategies (i.e., offline computation and reusability of intermediate results) into 
consideration, we could optimize the performance of PPODC. Without loss of generality, 
let us denote such an implementation by \ppodcopt. 
We estimated   
the online and offline computational costs, measured 
in terms of required multiplication (mul.) or exponentiation (exp.) operations, for each stage of \ppodcopt~separately. 
The results are given in Table \ref{tb:computation-analysis}. Here $m$ denotes 
the sum of the data records of all users.  
It is important to note that Stage 1 of \ppodcopt~is run only once whereas Stages 2 
and 3 are run in an iterative fashion until the termination condition holds.


The total communication costs for each stage of 
\ppodcopt~are extensively analyzed and the results are shown in Table \ref{tb:comm-analysis}. 
Here $K$ denotes the size (in bits) of the Paillier encryption key \cite{paillier-99}. 
Following from our analyses, we can observe 
that the costs (both computation and communication) of Stage 2 are 
significantly higher (depends on $m$) than the costs of Stage 3 in each iteration.

\section{Experimental Results}\label{sec:exp} 
First of all, we emphasize that PPODC is 100\% accurate in the sense that the outputs returned by PPODC 
and the standard $k$-means clustering algorithm (applied on the corresponding plaintext data) are the same. 
Therefore, in this section, we extensively analyze the computation costs of 
PPODC by performing various experiments using a real dataset under different parameter 
settings. Note that ours is the first work to address the PPODC problem and thus there 
exist no prior work to compare with our protocol.  

\subsection{Platform and Dataset Description} 
We implemented the protocols (both the direct implementation and optimized version of PPODC) in C 
using the GNU Multiple Precision Arithmetic (GMP) library\cite{gmp-lib}. For the optimized version of PPODC (denoted by \ppodcopt), we considered both the performance 
improvement strategies mentioned in Section \ref{sec:perf-anal}. The experiments were 
conducted on two Linux machines (playing the roles of $C_1$ and $C_2$), each with an 
Intel\textregistered~Core\texttrademark~i7-2600 CPU (3.40GHz) and 8GB RAM, running Linux version 3.12.6. The 
two machines were communicating over a TCP/IP network. 

For our experiments on real dataset, we used the KEGG Metabolic Reaction Network (Undirected) dataset 
from the UCI KDD archive \cite{ucikdd-kegg} that consists of 65,554 data records and 29 attributes. 
Since some of the attribute values are 
missing in the dataset, we removed the corresponding 
data records and the resulting dataset consists of 64,608 data records. As part of 
the pre-processing, we normalized the attribute values and scaled them into the 
integer domain $[0, 1000]$. Then we selected sample datasets (from the preprocessed data) 
by choosing data records at random based on the parameter 
values under consideration. We fixed the Paillier encryption key size to 1,024 bits 
(a commonly accepted key size) in all our experiments. 
For each sample dataset, we encrypted each of its data record attribute-wise 
using the Paillier encryption function \cite{paillier-99} and stored this encrypted data 
on the first machine. Note that the corresponding secret key $sk$ is stored on the second machine. 

We executed PPODC and \ppodcopt~over the encrypted data stored in the first machine under the above setting. 
The results presented in the rest of this section are averaged over ten sample datasets. 


\subsection{Empirical Analysis using Real Dataset}
To see the actual efficiency gains of \ppodcopt~over PPODC, we first 
evaluated their computation costs using different sampled datasets 
of varying sizes. Specifically, we fix the value of $l$ and $k$ to 10 and 8, respectively, and 
executed PPODC and \ppodcopt~on datasets of varying number of records $m$. The 
results per iteration are shown in Table \ref{tb:comparison-ppodc}. On the one hand, 
the running time of PPODC varies from 31.88 to 159.4 minutes when $m$ varies from 2,000 
to 10,000. On the other hand, the online running time of \ppodcopt~varies 
from 11.72 to 58.58 minutes when $m$ varies from 2,000 to 10,000. From these results, 
it is clear that the online computation time of the optimized version of PPODC is around 
2.7 times less than the online computation time of the direct implementation of PPODC. That is,   
the performance improvement strategies proposed in Section \ref{sec:perf-anal} 
boost the performance of PPODC by 60-65\%. We emphasize that 
the running time reported in this section also includes the communication costs, such 
as packet encoding and decoding, and network delays. 

\begin{table}[!t]
\centering
\renewcommand{\arraystretch}{1.5}
\setlength{\tabcolsep}{6pt}
\caption{Comparison of running time (in mins) for $l=10$ and $k=8$}
\begin{tabular}{|c|c|c|c|}
\hline
$\textbf{\emph m}$ & $\textbf{PPODC}$ & \multicolumn{2}{|c|}{\textbf{\ppodcopt}} \\
\cline{3-4}
 & (Direct Implementation) & (Online + Offline) & (Online)\\
\hline
2,000 & 31.88  & 23.52 & 11.72\\
\hline
4,000 & 63.76 & 47.04 & 23.43\\
\hline
6,000 & 95.64 & 70.56  & 35.15\\
\hline
8,000 & 127.52 & 94.08 & 46.87\\
\hline
10,000 & 159.4 & 117.6 & 58.58 \\
\hline
\end{tabular}
\label{tb:comparison-ppodc}
\end{table}

Having shown the performance improvement of \ppodcopt~over PPODC, we next 
analyze the online computation costs of \ppodcopt~based on different parameters. 
The computation cost of \ppodcopt~per iteration mainly depends on three  
parameters: (i) the number of data records of all users ($m$), (ii) 
the number of attributes ($l$), and (iii) the number of clusters ($k$). 
Therefore, we evaluate the performance of \ppodcopt~by varying these three parameters. 

\begin{figure*}[!t]
\centering
\subfigure[Running time for $m=6,000$]
{
     \epsfig{file=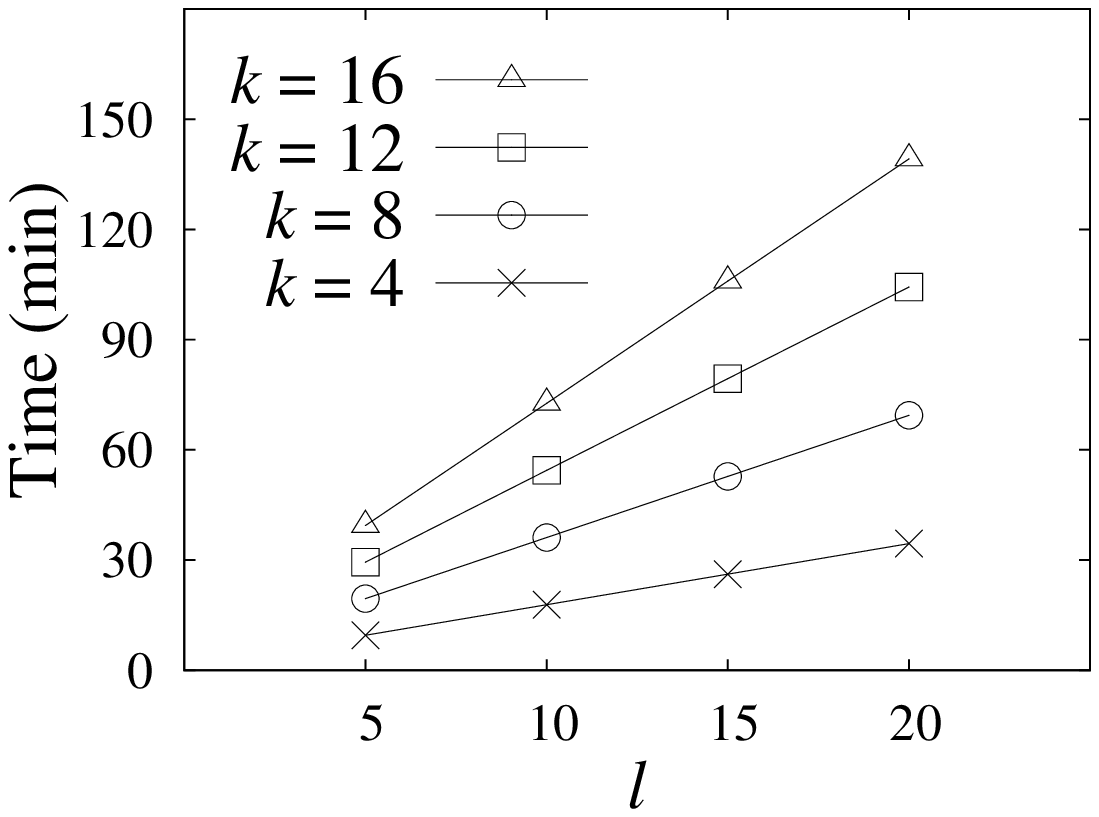, width= .29\textwidth}
\label{fig:figure2a}
}
\subfigure[Running time for $l=10$]
{
     \epsfig{file=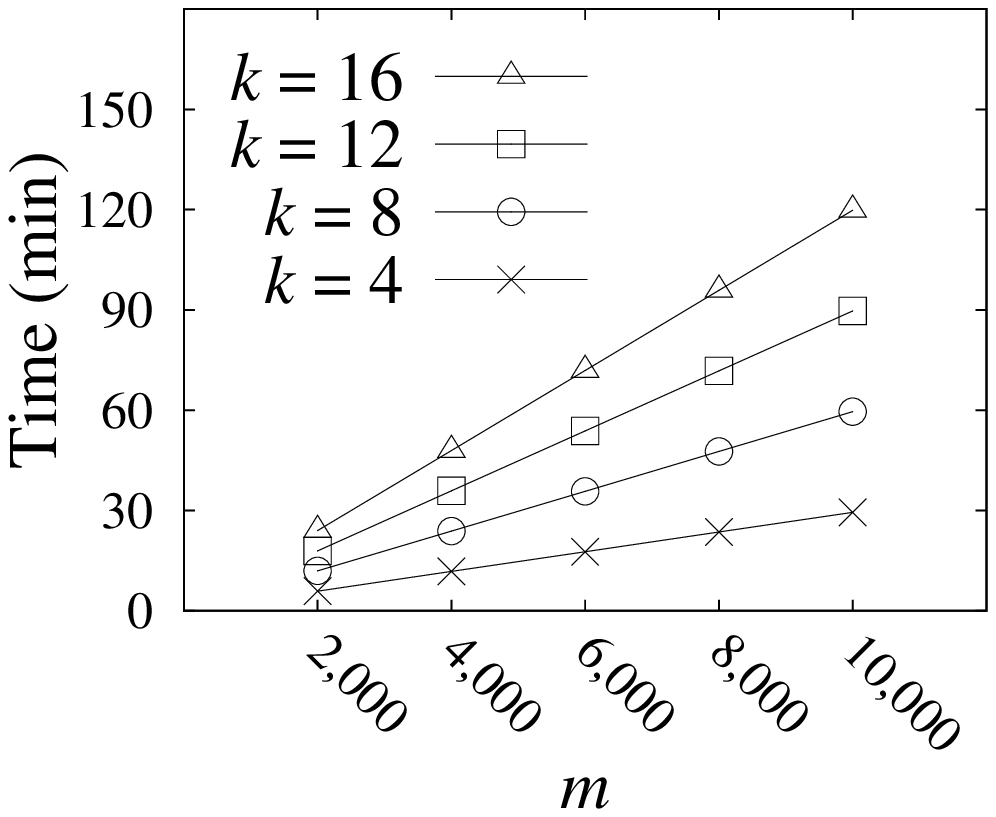, width= .29\textwidth}
\label{fig:figure2b}
}
\subfigure[Running time for $k=8$]
{
     \epsfig{file=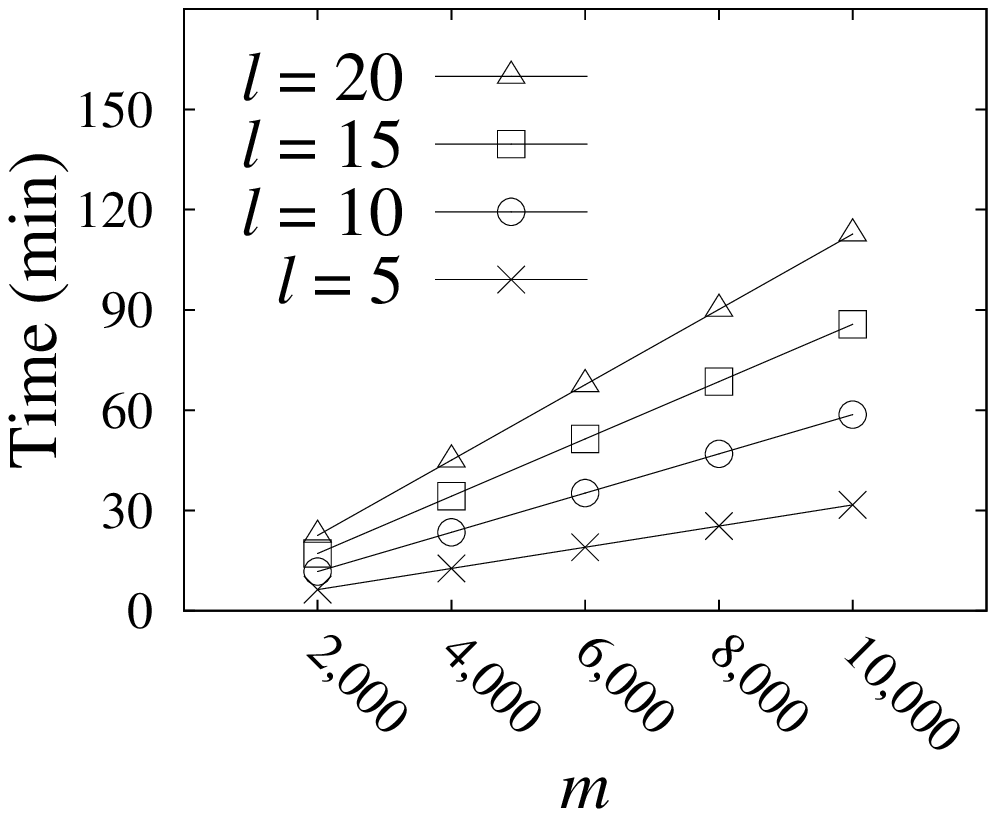, width= .29\textwidth}
\label{fig:figure2c}
}
\caption{Online computation costs of \ppodcopt~for encryption key size 1024 bits and varying values of $l, k,$ and $m$}
\vspace*{-0.6cm}
\end{figure*}

For $m=6,000$, Figure \ref{fig:figure2a} shows the online running time of \ppodcopt~for 
varying values of $l$ and $k$. For example, when $l=10$ and $k=8$, the online running 
time of \ppodcopt~is 36.14 minutes. The online running time of \ppodcopt~for $l=10$
and varying values of $k$ and $m$ are shown in Figure \ref{fig:figure2b}. The observation 
is that the running time grows linearly with $k$ and $m$. 
As shown in Figure \ref{fig:figure2c}, when $k=8$, a similar 
trend is observed for varying values of $m$ and $l$ . 
Putting everything together, it is clear that the running time 
of \ppodcopt~grows linearly with $m$, $k$ and $l$. This further justifies 
our theoretical analysis in Section \ref{sec:perf-anal}.

We observed that around 99\% of the computation 
time of \ppodcopt~is due to Stage 2. Also, the running time 
of each user is in few milliseconds (since he/she doesn't involve in any expensive operations),  
which makes our protocol very efficient 
from the end-user's computational perspective. 
In summary, the above results show that the proposed 
PPODC protocol, together with our optimizations, achieves 
reasonable efficiency given the stronger privacy guarantees.

\textbf{A Note on Scalability. } We emphasize that the computation costs of \ppodcopt~can be high for large datasets. 
However, it is worth noting that the performance of \ppodcopt~can be further improved by parallelizing 
the underlying operations. This is because the assignment of each data 
record to a new cluster in Stage 2 is independent of other records and thus we can almost parallelize 
the computations of Stage 2 at the record level. More specifically, $C_1$ and $C_2$ can 
utilize a cluster of nodes to perform their respective computations in parallel. Note that most of 
the current cloud service providers, such as Google and Amazon, typically support parallel 
processing on high performance computing nodes. Some 
of the large-scale parallel processing frameworks include Spark and Hadoop. 
Hence, by properly exploiting the parallel processing capability of clouds, we believe that the scalability issue 
of \ppodcopt~can be addressed to a great extent. 


\section{conclusions}\label{sec:future}
Existing privacy-preserving distributed clustering techniques, which  
can allow the users to collaboratively and securely perform the clustering task, 
incur heavy costs (both communication- and computation-wise) on the participating 
users. To address this issue, in this paper, we introduced the problem of privacy-preserving and outsourced 
distributed clustering (PPODC) where a set of users can securely outsource their databases  
and the intended clustering task to a cloud environment. 
We proposed a novel PPODC protocol under a federated cloud environment that 
can perform the $k$-means clustering on the users aggregated encrypted data in 
a privacy-preserving manner. At the core of our protocol, we proposed new  
transformations to construct an 
order-preserving Euclidean distance function and evaluate the 
termination condition of the $k$-means clustering algorithm over encrypted data. 

The proposed PPODC protocol ensures data confidentiality of all users and incurs 
negligible costs on the user side. We theoretically 
estimated the complexities of our protocol and 
experimentally analyzed its efficiency 
using a real dataset. Our results show 
that our protocol 
incurs reasonable costs on the cloud side and is practical for non-real-time 
applications. 
One important contribution of our protocol is that most of its underlying computations 
can be parallelized.  
As future work, we plan to implement the proposed protocol 
using parallelism on a cluster of nodes and evaluate its 
performance. 
Also, we will extend the research ideas proposed in this paper to 
other data mining tasks, such as classification, association rule mining, and 
regression analysis.

\bibliographystyle{IEEEtran}
\bibliography{ref}  

\begin{thebibliography}{10}
\providecommand{\url}[1]{#1}
\csname url@samestyle\endcsname
\providecommand{\newblock}{\relax}
\providecommand{\bibinfo}[2]{#2}
\providecommand{\BIBentrySTDinterwordspacing}{\spaceskip=0pt\relax}
\providecommand{\BIBentryALTinterwordstretchfactor}{4}
\providecommand{\BIBentryALTinterwordspacing}{\spaceskip=\fontdimen2\font plus
\BIBentryALTinterwordstretchfactor\fontdimen3\font minus
  \fontdimen4\font\relax}
\providecommand{\BIBforeignlanguage}[2]{{%
\expandafter\ifx\csname l@#1\endcsname\relax
\typeout{** WARNING: IEEEtran.bst: No hyphenation pattern has been}%
\typeout{** loaded for the language `#1'. Using the pattern for}%
\typeout{** the default language instead.}%
\else
\language=\csname l@#1\endcsname
\fi
#2}}
\providecommand{\BIBdecl}{\relax}
\BIBdecl

\bibitem{data-cluster1}
A.~K. Jain, M.~N. Murty, and P.~J. Flynn, ``Data clustering: a review,''
  \emph{ACM Computing Surveys}, vol.~31, pp. 264--323, September 1999.

\bibitem{data-cluster2}
P.~Berkhin, ``A survey on clustering data mining techniques,'' in \emph{In:
  Grouping Multidimensional Data}.\hskip 1em plus 0.5em minus 0.4em\relax
  Springer, 2006, pp. 25--71.

\bibitem{data-cluster3}
M.~A. Dalal and N.~D. Harale, ``A survey on clustering in data mining,'' in
  \emph{Proceedings of the International Conference \& Workshop on Emerging
  Trends in Technology}.\hskip 1em plus 0.5em minus 0.4em\relax ACM, 2011, pp.
  559--562.

\bibitem{ir-cluster}
P.~Patrick and L.~Dekang, ``Document clustering with committees,'' in
  \emph{SIGIR}.\hskip 1em plus 0.5em minus 0.4em\relax ACM, 2002, pp. 199--206.

\bibitem{ml-cluster}
R.~Michalski and R.~Stepp, \emph{In Machine Learning: An Artificial
  Intelligence Approach}.\hskip 1em plus 0.5em minus 0.4em\relax Tiago
  Publishing Co., 1983, ch. Learning from Observation: Conceptual Clustering,
  pp. 331--363.

\bibitem{pr-cluster}
B.~Andrea and B.~Palma, ``A survey of fuzzy clustering algorithms for pattern
  recognition,'' \emph{IEEE Transactions on Systems, Man, and Cybernetics},
  vol.~29, no.~6, pp. 778--785, December 1999.

\bibitem{ia-cluster}
A.~Jain and P.~flynn, \emph{In Advances in Image Understanding: A Festschrift
  for Azriel Rosenfeld}.\hskip 1em plus 0.5em minus 0.4em\relax IEEE Press,
  1996, ch. Image Segmentation using Clustering, pp. 65--83.

\bibitem{tm-cluster}
B.~Michael and C.~Malu, \emph{Survey of Text Mining II: Clustering,
  Classification, and Retrieval}.\hskip 1em plus 0.5em minus 0.4em\relax
  Springer, 2007.

\bibitem{pp-kmeans1}
J.~Vaidya and C.~Clifton, ``Privacy-preserving k-means clustering over
  vertically partitioned data,'' in \emph{ACM SIGKDD}, 2003, pp. 206--215.

\bibitem{ppdc-agg-cluster}
C.~Su, J.~Zhou, F.~Bao, T.~Takagi, and K.~Sakurai, ``Two-party
  privacy-preserving agglomerative document clustering,'' in
  \emph{ISPEC}.\hskip 1em plus 0.5em minus 0.4em\relax Springer-Verlag, 2007,
  pp. 193 -- 208.

\bibitem{pp-kmeans2}
G.~Jagannathan and R.~Wright, ``Privacy-preserving distributed k-means
  clustering over arbitrarily partitioned data,'' in \emph{ACM SIGKDD}, 2005,
  pp. 593--599.

\bibitem{bunn-2007}
P.~Bunn and R.~Ostrovsky, ``Secure two-party k-means clustering,'' in \emph{ACM
  CCS}, 2007, pp. 486--497.

\bibitem{gentry-09}
C.~Gentry, ``Fully homomorphic encryption using ideal lattices,'' in \emph{ACM
  STOC}, 2009, pp. 169--178.

\bibitem{gentry-2011}
C.~Gentry and S.~Halevi, ``Implementing gentry's fully-homomorphic encryption
  scheme,'' in \emph{EUROCRYPT}.\hskip 1em plus 0.5em minus 0.4em\relax
  Springer, 2011, pp. 129--148.

\bibitem{lloyd-1982}
S.~Lloyd, ``Least squares quantization in pcm,'' \emph{IEEE Transactions on
  Information Theory}, vol.~28, no.~2, pp. 129--137, 1982.

\bibitem{fukunaga-1990}
K.~Fukunaga, \emph{Introduction to Statistical Pattern Recognition (2Nd
  Ed.)}.\hskip 1em plus 0.5em minus 0.4em\relax San Diego, CA, USA: Academic
  Press Professional, Inc., 1990.

\bibitem{nist-federated-cloud}
{NIST}, ``Nist us government cloud computing technology roadma,'' Volume I:
  High Priority Requirements to Further USG Agency Cloud Computing Adoption,
  {November 2011. Special Publication. 500-293},
  \url{http://www.nist.gov/itl/cloud/upload/SP_500_293_volumeI-2.pdf}.

\bibitem{goldreich-book-gcp}
O.~Goldreich, \emph{The Foundations of Cryptography}.\hskip 1em plus 0.5em
  minus 0.4em\relax Cambridge University Press, 2004, vol.~2, ch. General
  Cryptographic Protocols.

\bibitem{paillier-99}
P.~Paillier, ``Public key cryptosystems based on composite degree residuosity
  classes,'' in \emph{Eurocrypt}.\hskip 1em plus 0.5em minus 0.4em\relax
  Springer-Verlag, 1999, pp. 223--238.

\bibitem{damgard-tpc-2001}
I.~Damg{\aa}rd and M.~Jurik, ``A generalisation, a simplification and some
  applications of paillier's probabilistic public-key system,'' in
  \emph{PKC}.\hskip 1em plus 0.5em minus 0.4em\relax Springer-Verlag, 2001, pp.
  119--136.

\bibitem{hazay-2012}
C.~Hazay, G.~L. Mikkelsen, T.~Rabin, and T.~Toft, ``Efficient rsa key
  generation and threshold paillier in the two-party setting,'' in
  \emph{CT-RSA}.\hskip 1em plus 0.5em minus 0.4em\relax Springer-Verlag, 2012,
  pp. 313--331.

\bibitem{ivan-2003}
A.-A. Ivan and Y.~Dodis, ``Proxy cryptography revisited,'' in \emph{NDSS},
  2003.

\bibitem{ateniese-2006}
G.~Ateniese, K.~Fu, M.~Green, and S.~Hohenberger, ``Improved proxy
  re-encryption schemes with applications to secure distributed storage,''
  \emph{ACM TISSEC}, vol.~9, no.~1, pp. 1--30, Feb. 2006.

\bibitem{liu-2014}
D.~Liu, E.~Bertino, and X.~Yi, ``Privacy of outsourced k-means clustering,'' in
  \emph{ACM ASIACCS}, 2014, pp. 123--134.

\bibitem{ppdm1}
R.~Agrawal and R.~Srikant, ``Privacy preserving data mining,'' in \emph{ACM
  SIGMOD}, vol.~29, 2000, pp. 439--450.

\bibitem{ppdm2}
Y.~Lindell and B.~Pinkas, ``Privacy preserving data mining,'' in \emph{Journal
  of Cryptology}, vol.~15, 2002, pp. 177 -- 206.

\bibitem{upmanyu-2010}
M.~Upmanyu, A.~Namboodiri, K.~Srinathan, and C.~Jawahar, ``Efficient privacy
  preserving k-means clustering,'' in \emph{Intelligence and Security
  Informatics}.\hskip 1em plus 0.5em minus 0.4em\relax Springer, 2010, vol.
  6122, pp. 154--166.

\bibitem{goldwasser-1989}
S.~Goldwasser, S.~Micali, and C.~Rackoff, ``The knowledge complexity of
  interactive proof systems,'' \emph{SIAM Journal on Computing}, vol.~18,
  no.~1, pp. 186--208, Feb. 1989.

\bibitem{yousef-icde14}
Y.~Elmehdwi, B.~K. Samanthula, and W.~Jiang, ``Secure k-nearest neighbor query
  over encrypted data in outsourced environments,'' in \emph{ICDE}.\hskip 1em
  plus 0.5em minus 0.4em\relax IEEE, 2014, pp. 664--675.

\bibitem{bksam-asiaccs13}
B.~K. Samanthula, C.~Hu, and W.~Jiang, ``An efficient and probabilistic secure
  bit-decomposition,'' in \emph{8th {ACM} Symposium on Information, Computer
  and Communications Security, {ASIACCS}}, 2013, pp. 541--546.

\bibitem{nishide-2007}
T.~Nishide and K.~Ohta, ``Multiparty computation for interval, equality, and
  comparison without bit-decomposition protocol,'' in \emph{Proceedings of the
  10th International Conference on Practice and Theory in Public-key
  Cryptography}, ser. PKC'07.\hskip 1em plus 0.5em minus 0.4em\relax Berlin,
  Heidelberg: Springer-Verlag, 2007, pp. 343--360.

\bibitem{Shamir79}
A.~Shamir, ``How to share a secret,'' \emph{Communications of the ACM},
  vol.~22, no.~11, pp. 612 -- 613, November 1979.

\bibitem{Goldreichenc}
\BIBentryALTinterwordspacing
O.~Goldreich, \emph{The Foundations of Cryptography}.\hskip 1em plus 0.5em
  minus 0.4em\relax Cambridge University Press, 2004, vol.~2, ch. Encryption
  Schemes. [Online]. Available:
  \url{http://www.wisdom.weizmann.ac.il/~oded/PSBookFrag/enc.ps}
\BIBentrySTDinterwordspacing

\bibitem{bksam-ppknn-tech}
B.~K. Samanthula, Y.~Elmehdwi, and W.~Jiang, ``k-nearest neighbor
  classification over semantically secure encrypted relational data,'' eprint
  arXiv:1403.5001, 2014, \url{http://arxiv.org/abs/1403.5001}.

\bibitem{damgard-2010}
I.~Damgård, M.~Jurik, and J.~B. Nielsen, ``A generalization of paillier's
  public-key system with applications to electronic voting,''
  \emph{International Journal of Information Security}, vol.~9, no.~6, pp.
  371--385, Dec. 2010.

\bibitem{gmp-lib}
{The GNU MP Bignum Library}, \url{https://gmplib.org/}.

\bibitem{ucikdd-kegg}
M.~Naeem and S.~Asghar, ``{KEGG Metabolic Reaction Network (Undirected) Data
  Set},'' The UCI KDD Archive, 2011,
  \url{https://archive.ics.uci.edu/ml/datasets/KEGG+Metabolic+Reaction+Network%
+(Undirected)}.

\end{thebibliography}

\end{document}